\documentclass{aa}  
\RequirePackage[colorlinks=true
,urlcolor=blue
,anchorcolor=blue
,citecolor=blue
,filecolor=blue
,linkcolor=blue
,menucolor=blue
,linktocpage=true
,pdfproducer=medialab
,pdfa=true
]{hyperref}
\makeatletter
\renewcommand*\aa@pageof{, page \thepage{} of \pageref*{LastPage}}
\makeatother
\usepackage[binary-units, separate-uncertainty=true]{siunitx}
\DeclareSIUnit \mJy {mJy}
\DeclareSIUnit \GHz{GHz}
\DeclareSIUnit \MHz{MHz}
\usepackage[frozencache,cachedir=.]{minted}
\usepackage{natbib}
\usepackage{comment}
\usepackage{stackengine}
\newcommand\xrowht[2][0]{\addstackgap[.5\dimexpr#2\relax]{\vphantom{#1}}}
\bibpunct{(}{)}{;}{a}{}{,}
\usepackage[english]{babel}
\usepackage{amsmath}
\usepackage[utf8]{inputenc}
\usepackage[caption=false]{subfig}
\usepackage[T1]{fontenc}
\usepackage{epigraph}
\usepackage{xcolor}
\usepackage{colortbl}
\usepackage{import}
\usepackage{float}
\usepackage{array}
\usepackage{graphicx}
\usepackage{multicol}
\usepackage{cancel,soul,ulem}
\usepackage{listings}
\usepackage{txfonts}

\begin{document}

   \title{A cross-correlation analysis of CMB lensing and radio galaxy maps}

   \author{G. Piccirilli\inst{1,} \inst{2}\thanks{Email:giulia.piccirilli@roma2.infn.it} \and
   M. Migliaccio\inst{1,} \inst{2} \and E. Branchini\inst{3,} \inst{4} \and
   A. Dolfi\inst{5} 
          }

   \institute{Dipartimento di Fisica, Università di Roma Tor Vergata, via della Ricerca Scientifica, 1, 00133, Roma, Italy
   \and INFN - Sezione di Roma 2, Università di Roma Tor Vergata, via della Ricerca Scientifica, 1, 00133 Roma, Italy
   \and Department of Physics, University of Genova, Via Dodecaneso 33, 16146 Genova, Italy
   \and INFN - Sezione di Roma Tre, via della Vasca Navale 84, I-00146 Roma, Italy
   \and Centre for Astrophysics \& Supercomputing, Swinburne University of Technology, Hawthorn VIC 3122, Australia  }


\abstract{}{}{}{}{} 
  \abstract
   {}
   {
   The goal of this work is to clarify the origin of the seemingly anomalously large clustering signal detected at large angular separation in the wide TGSS radio survey and, in so doing, to investigate the nature and the clustering properties of the sources that populate the radio sky in the [0.15,1.4] GHz frequency range.  
   }
   {
   To achieve this goal, we cross-correlate the angular position of the radio sources in the TGSS and NVSS samples with the Cosmic Microwave Background (CMB) lensing maps from the Planck satellite. A cross-correlation between two different tracers of the underlying mass density field has the advantage to be quite insensitive to possible systematic errors that may affect the two observables, provided that these are not correlated, which seems unlikely in our case. The cross-correlation analysis is performed in harmonic space and limited to relatively modest multipoles. These choices, together with that of binning the measured spectra, minimize the correlation among the errors in the measured spectra and allow us to adopt the Gaussian hypothesis to perform the statistical analysis. Finally, we decide to consider the auto-angular power spectrum on top of the cross-spectrum since a joint analysis has the potential to improve the constraints on the radio sources properties by lifting the degeneracy between the redshift distribution $N(z)$ and the bias evolution $b(z)$.
   }
   {
   The angular cross-correlation analysis
   does not present the power excess at large scales for TGSS and provides a TGSS - CMB lensing cross-spectrum which is in agreement with the one measured using the NVSS catalog. This result strongly suggests that the excess found in TGSS clustering analyses can be due to uncorrected systematic effects in the data. However, we consider several cross-spectra models that rely on physically motivated combinations of $N(z)$ and $b(z)$ prescriptions for the radio sources and find that they all underestimate the amplitude of the measured cross-spectra on the largest angular scales of $\sim 10^{\circ}$ considered in our analysis. This result is robust to the various potential sources of systematic errors, both of observational and theoretical nature, that may affect our analysis, including the uncertainties in the $N(z)$ model. Having assessed the robustness of the results to the choice of $N(z)$, we repeat the analysis using simpler bias models specified by a single free parameter $b_g$, namely the value of the effective bias of the radio sources at redshift zero. This improves the goodness of the fit although not even the best model, which assumes a non evolving bias, quite matches the amplitude of the cross-spectrum at small multipoles. Moreover, the best fitting bias parameter $b_g=2.53\pm 0.11$ appears to be somewhat large considering that it represents the effective bias of a sample that is locally dominated by mildly clustered star forming galaxies and Fanaroff-Riley class I sources. Interestingly, is the addition of the angular auto-spectrum that favors the constant bias model over the evolving one.
   }
   {
   The nature of the large cross-correlation signal between the radio sources and the CMB lensing maps found in our analysis at large angular scales is not clear. It probably indicates some limitation in the modeling of the radio sources, i.e. the relative abundance of the various populations, their clustering properties and how these evolve with redshift. What our analysis does show is the importance of combining the auto-spectrum with the cross-spectrum, preferably obtained with unbiased tracers of the large scale structure, like the CMB lensing, to answer these questions.  
   }

   \maketitle
\section{Introduction}
\label{sec:intro}

The study of the cosmic Large Scale Structure (LSS) has the potential to shed light on the nature of the `dark' physics, that drives the accelerated expansion of the Universe, or to detect deviations from the General Relativity framework. 
Among the many available LSS probes, extra-galactic radio sources possess two important properties: they are bright and they are less affected by Galactic dust extinction than optical sources. As a result, radio sources can trace the spatial distribution of matter over a large volume of the Universe.

For this reason, radio sources have been extensively exploited to study the LSS of the Universe. Clustering analyses have been performed using wide surveys like FIRST \citep{cress}, 87GB \citep{GB87} and PMN \citep{loan}, WENSS \citep{rengelink}, SUMSS \citep{blake04}, NVSS \citep{Blake_Wall_2002,overzier,negrello,chen16}, TGSS-ADR1 \cite{rana} and LoTSS \citep{LoTSS1}. Those analyses used the angular two-point correlation function to characterize their clustering properties.
Similar studies have been performed in harmonics space to estimate the angular power spectrum of the sources in the NVSS \citep{blake04,Nusser_2015}, TGSS-ADR \citep{dolfi_2019,Tiwari_2019} and LoTSS \citep{tiwari21} catalogs.

Clustering analyses of radio objects have produced a number of intriguing results. The most remarkable one is the amplitude of their angular dipole that turned out to be significantly larger than $\Lambda$CDM expectations \citealt{singal,Giblelyou,rubart,fernandez,tiwari15,tiwari16,colin17,bengaly18,siewert21}. 
Anomalously large clustering power has also been detected at low multipoles in the TGSS-ADR survey \citep{dolfi_2019,Tiwari_2019} but not in the NVSS catalog \citep{tiwari19_2,dolfi_2019}.
A further anomaly is the amplitude of anisotropy measured in the Radio synchrotron background at 140 MHz which is also larger than expected \citep{offringa}. These results can be interpreted as either a challenge to the standard cosmological model or the manifestation of observational systematics that have not been properly accounted for.
One example of this second case is given by systematic uncertainties in the flux calibration  that can potentially generate spurious clustering signal on large angular scales. Systematic errors of this type have been identified in both the TGSS-ADR \citep{Tiwari_2019} and in the LoTSS-DR1 data-sets \citep{tiwari21} and may help in relieving the tension with the $\Lambda$CDM model.

Performing a uniform flux calibration is notoriously challenging for all type of wide surveys. Those in the radio bands present two additional problems.
The first one is represented by the composite nature of the radio sources that include local, relatively faint, star forming galaxies as well as bright and distant radio quasars. They map the underlying mass distribution differently, i.e. they are characterized by different biasing relations and, therefore, possess different clustering properties. Moreover, their relative abundance depends on the redshift and on the radio frequency. Therefore, both the effective biasing and the redshift distribution of the radio sources depend on the survey characteristics and need to be modeled accordingly.
The second problem is the small fraction of radio objects with measured spectroscopic redshift which makes it impossible to map their spatial distribution and limits clustering analyses to angular two-point statistics. The two point auto-correlation signal (or equivalently, the auto angular power spectrum) of the radio sources depends on the combined effect of biasing relation and redshift distribution, so that these two functions cannot be estimated independently. One way to alleviate, at least in part, these problems is to cross-correlate the angular distribution of the radio sources with that of a different type of mass tracer. Since observational systematics (like the aforementioned flux calibration issues) and foreground emissions are not expected to correlate with the LSS, spurious angular clustering signals that may contribute to the auto-correlation signal will not affect the cross-correlation one. Moreover, if the second catalog contains objects with a known bias and redshift distribution or, better still, if it is an unbiased tracer of the mass field, then it will be possible to infer both the bias model and the redshift distribution of the radio sources.

The gravitational lensing signal of the background objects does trace the underlying mass in an unbiased way, and therefore its maps are ideal to cross-correlate catalogs of radio sources with.
However, lensing maps, obtained from the images of carefully selected background galaxies (see e.g. \citealt{kids,des}), do not cover the same wide areas as the radio surveys. For this reason, we shall consider instead the all-sky  gravitational lensing maps of the Cosmic Microwave Background (CMB) photons. Cross-correlation analyses of CMB lensing and NVSS radio sources have been already successfully performed, leading to a high significance detection of the cross-correlation signal \citep{Smith_2007,hirata08,feng,planck_lens2013, giannantonio2013}. Our scope is to expand these studies to other catalogs of radio sources with the aim of inferring their nature, distribution and clustering properties.

More in detail, our goal is twofold. First of all, we cross-correlate the angular positions of the TGSS-ADR radio sources with the Planck lensing maps \citep{planck_lens}
to clarify the nature of the large scale clustering excess seen in the auto-correlation analysis. The detection of a similar excess in the cross-correlation would point at a possible cosmological origin.
Secondly, we compare the measured cross-correlation between NVSS, TGSS and the CMB lensing maps with theoretical predictions to constrain the biasing function and the redshift distribution of the radio sources, also including the auto-correlation statistics in the analysis. The analysis we perform in this work shares the same goals as the one recently performed by \citep{Alonso_2021} using a newer data-set (the LoTSS radio survey) distributed over a much smaller area than the one we consider here. A similar analysis was also performed using the cross-spectrum between the CMB lensing from ACT (Atacama Cosmology Telescope) and the FIRST catalog, leading to constraints on the bias and the typical halo mass of radio-loud AGNs \citep{Allison_2015}. 

The layout of the paper is as follows. In Section \ref{sec:data} we present the data-sets (CMB-lensing maps and radio catalogs) used in this work. In Section \ref{sec:theory} we discuss the theoretical model for the auto and the cross angular spectra that we measured from the data using the estimators presented in Section \ref{sec:estimators}. The results of the measured spectra, their comparison with theoretical predictions and the inferred quantities are presented in Section~\ref{sec:modelvsdata}. In Section~\ref{sec:bias_fit}, we estimated the galaxy bias parameter while keeping the redshift distribution model fixed and in Section~\ref{sec:robustness_tests} we explored different tests, to assess the robustness of cross-correlation analysis against possible systematics in the radio catalogs, in the lensing map or both. Finally, in  Section~\ref{sec:conclusion} we discuss our results and present our main conclusions.\\
Throughout this paper we assume a flat $\Lambda CDM$ cosmological model characterized by the cosmological parameters of \cite{planck_params}.
\section{Data-sets}
\label{sec:data}
In this section, we briefly describe the data-sets considered for our analysis. The CMB lensing maps and the catalogs of radio sources are respectively presented in Section~\ref{sec:lensing_data} and in Section~\ref{sec:radio_cat}.
\subsection{CMB lensing convergence map: Planck data}
\label{sec:lensing_data}
The Planck satellite has observed the sky in several microwave bands
with an unprecedented sensitivity and angular resolution. 
One of its main scientific achievements is the high precision measurement of the gravitational lensing effect, with the reconstruction of the lensing potential map over $67\%$ of the sky and the measurement of its angular power spectrum in the multipole range $8\leq \ell \leq 2000$ (\citealt{planck_lens}). In this paper, we  use one of the maps produced by the Planck collaboration, namely the CMB convergence one. This map was obtained with a minimum variance (MV) estimator that combines the reconstructions from CMB temperature and polarization anisotropies, thanks to which the lensing signal has been detected with a $40 \sigma$ significance.
We use the spherical harmonic coefficients for the lensing convergence provided by the Planck collaboration\footnote{\url{https://wiki.cosmos.esa.int/planck-legacy-archive/index.php/Lensing}} to generate a \mintinline{Python}{HEALPix}\footnote{\url{https://healpix.sourceforge.io/}}(\citealt{Gorski_2005}) map with a resolution parameter Nside = $512$. This choice corresponds to an angular resolution of $\sim 7$ arcmin and it is motivated by the fact that we expect negligible contributions to the cross-correlation from scales smaller than these.
The Planck collaboration also provides the noise spectrum of the convergence map, $N_{\ell}^{\kappa \kappa}$, together with a binary map to mask out the sky area to be excluded from the lensing analysis because of possible residual contamination by Galactic and extra-galactic foregrounds (see Figure~\ref{fig:maps_radio}).\\
\subsection {Radio catalogs}\label{sec:radio_cat}
\begin{figure}[!ht]
\centering
\includegraphics[width=\hsize]{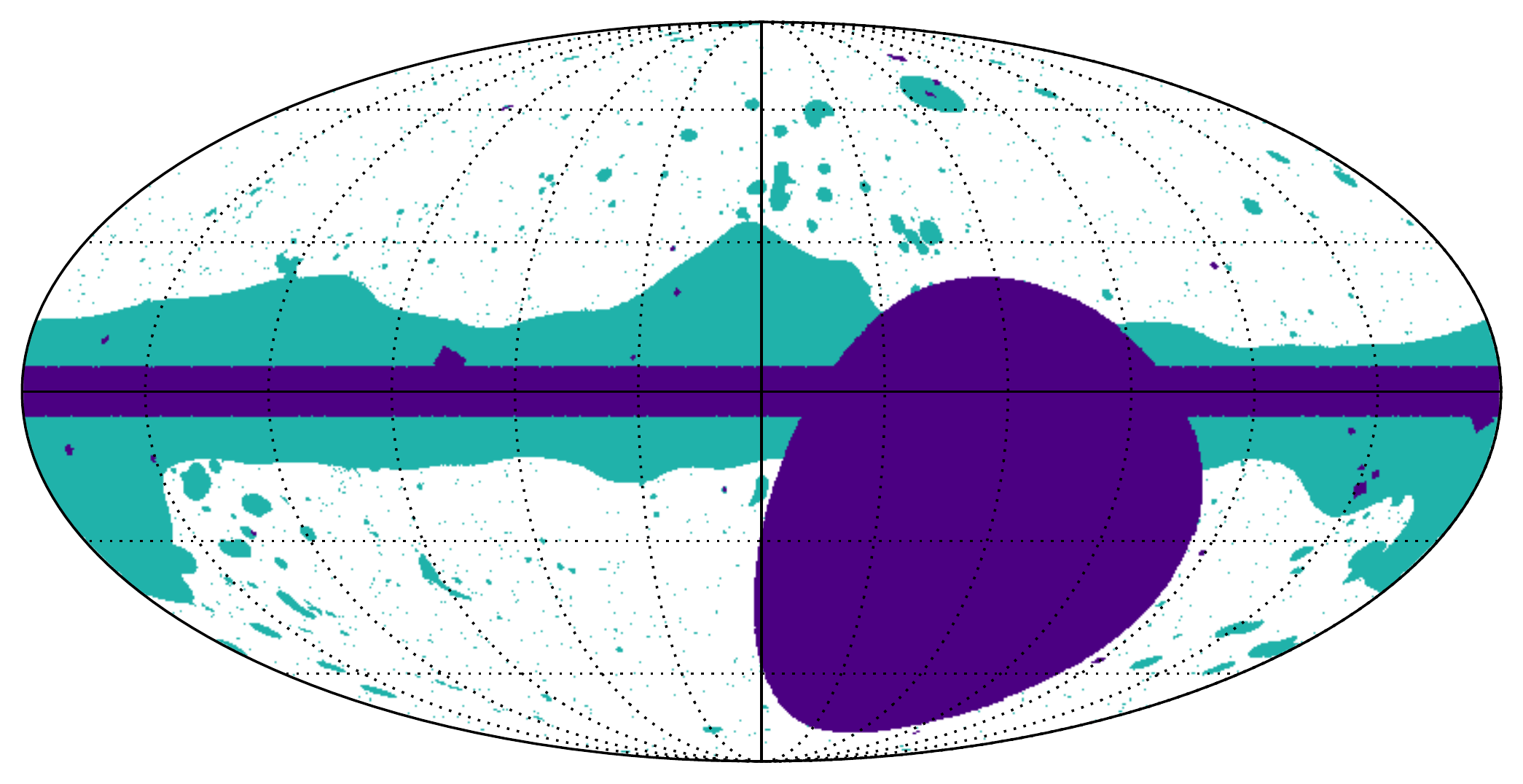}
\includegraphics[width=\hsize]{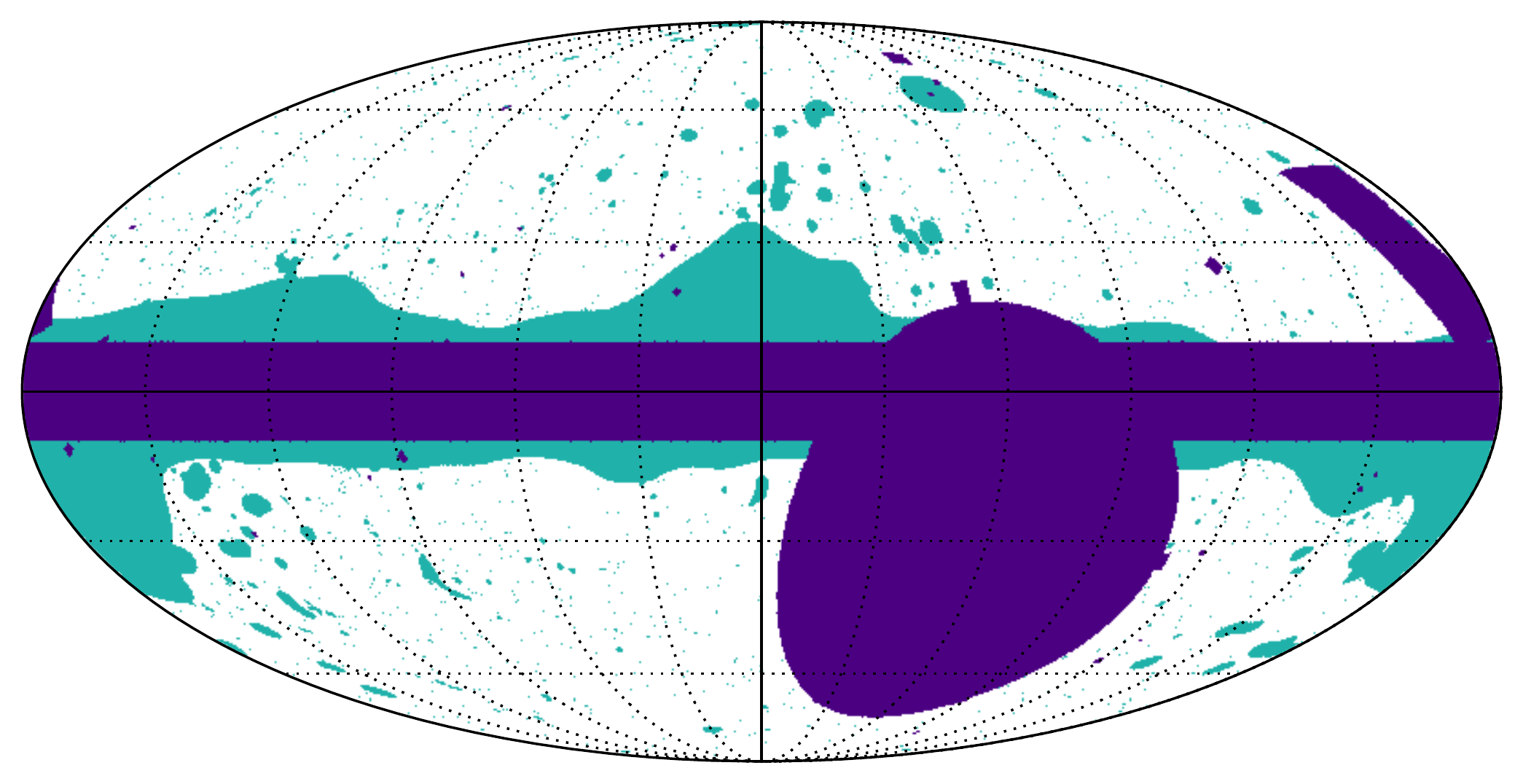}
\caption{Mollweide projection in Galactic coordinates of the masked areas in the radio samples (purple) and in the Planck CMB lensing convergence map (turquoise). Top panel: NVSS case. Bottom panel: TGSS case.}
\label{fig:maps_radio}
\end{figure}

The second data-set that we consider here consists of two catalogs of extra-galactic radio sources derived from the NRAO VLA Sky Survey at  $\SI{1.4}{\GHz}$ (NVSS hereafter, \citealt{Condon_1998}) and from the TIFR GMRT Sky Survey at $\SI{150}{\MHz}$ (TGSS, \citealt{intema}).\\
The NVSS catalog was obtained using the Very Large Array Telescope (VLA\footnote{\url{https://public.nrao.edu/telescopes/vla/}}) between 1993 and 1996; while TGSS data have been collected at the Giant Metrewave Radio Telescope\footnote{\url{http://www.gmrt.ncra.tifr.res.in/}} radio telescope between 2010 and 2012. 
The two data-sets of radio sources are distributed over wide and largely overlapping areas, $ \Delta \Omega $, covering a fraction of the sky, $f_{sky} \equiv \Delta \Omega /4\pi$, excluding regions in the Southern Hemisphere (with declination $\delta<-40^{\circ}$ and $\delta<-45^{\circ}$ for NVSS and TGSS, respectively) and with low Galactic latitudes ($|b|<5^{\circ}$ and $|b|<10^{\circ}$ again for NVSS and TGSS). Moreover, following \citealt{intema}, we have removed TGSS objects in the area $25^{\circ}<\delta<39^{\circ}$, $97.5^{\circ}<\alpha<142.5^{\circ}$ that were observed in bad ionospheric conditions.
Finally, in both samples, the regions around the brightest objects have also been masked out (\citealt{Nusser_2015}).
In Figure~\ref{fig:maps_radio}, we show the portion of masked sky in both TGSS and NVSS catalogs (purple regions) together with the masked area of CMB convergence map (turquoise areas). \\
After excluding radio objects in these regions, we end up with $109\ 940$ TGSS sources with fluxes in the interval $S_{150} = [200, 1000] \ \SI{}{\mJy}$ and with $518\ 894$ NVSS sources in the flux range $S_{1.4}=[10,1000] \ \SI{}{\mJy}$. The lower limit for the NVSS flux is related to the fact that spurious fluctuations of the surface density of the radio sources have been detected for objects fainter than this value \citep{blake_2004b}. TGSS sources are generally brighter than the NVSS ones and a large fraction of them have a counterpart in the NVSS catalog.
The main characteristics of the two catalogs are summarized in Table~\ref{tab:tgss_nvss}.\\
We use the angular position of these radio sources to build \mintinline{Python}{HEALPix} maps of objects counts with a resolution parameter Nside = $512$ matching the one of the CMB lensing convergence map. 
We define fluctuations in the objects number counts as:
\begin{equation}
\label{eq:deltaN}
    \delta_g(\hat{n}) = \frac{N(\hat{n})-\bar{N}}{\bar{N}},
\end{equation}
where $\hat{n}$ is the direction to the pixel, $N(\hat{n})$ is the number of radio objects in the pixel and $\bar{N}$ is their mean number per pixel.
\vspace{1.cm}
\begin{table}
\centering
\begin{tabular}{c c c c c c c}
\hline
\hline\xrowht[()]{5pt}
 catalog & $S^{min}$ & $S^{max}$ & $f_{sky}^{gg}$ &$f_{sky}^{\kappa g}$ & \# \\ [0.5ex] 
  & $[\SI{}{\mJy}]$ & $[\SI{}{\mJy}]$ & & & \\ [0.5ex]
 \hline\xrowht[()]{5pt}
 TGSS & 200 & 1000 & 0.70 & $0.57$ & 109\,940\\ [1ex] 
 NVSS & 10 & 1000 & 0.75 & $0.56$ & 518\,894\\ [0.5ex] 
 \hline
\end{tabular}
\caption{Main characteristics of the TGSS and NVSS reference samples adopted in this work. First column: sample name. Second and third columns: lower and upper flux cut in $[\SI{}{\mJy}]$. Fourth column: sky fraction covered by the sample. Fifth column: sky fraction in common between the radio sample and the CMB lensing convergence map. Sixth column: number of radio sources in the catalogs.}
\label{tab:tgss_nvss}
\end{table}
\section{Angular power spectra models}
\label{sec:theory}
CMB photons propagating from the last scattering surface are deflected by the gravitational potential of the LSS of the Universe. This phenomenon is known as weak gravitational lensing. It re-maps the CMB anisotropies leaving distinctive signatures in their distribution (\citealt{Lewis, Hanson_2010}), which can be used to reconstruct the lensing potential:
\begin{equation}
\label{eq:len_pot}
    \phi(\hat{n})=-2\int_{0}^{\infty} d\chi \frac{\chi^{\ast}-\chi}{\chi^{\ast}\chi}\Psi(\chi\hat{n},\eta_0-\chi),
\end{equation}
where $\Psi$ is the underlying gravitational potential, $\eta_0$ is the conformal time measured today and $\chi$ is the comoving distance. Note that this expression is valid for a flat Universe. All the quantities with the asterisk superscript are computed at the last scattering surface redshift, $z^*\simeq 1100$.

Starting from the lensing potential, we use the 2D Poisson equation to define the dimensionless lensing convergence: $\kappa(\hat{n})=-\frac{1}{2}\nabla^2\phi(\hat{n})$.
Since the gravitational potential depends on the matter overdensity $\delta_m$, we can express the CMB lensing convergence as (e.g. \citealt{Bianchini_2016}):
\begin{equation}
\kappa(\hat{n}) = \int_0^{\infty} dz W^{\kappa}(z)\delta_m(\chi(z)\hat{n},z).
\label{eq:kappa}
\end{equation}
In the above equation, $W^{\kappa}(z)$ is the convergence window function and its explicit dependence on the redshift is:
\begin{equation}
\label{eq:len_win}
W^{\kappa}(z)= \frac{3}{2c}\Omega_{m,0}\frac{H_0^2}{H(z)}(1+z)\chi(z) \frac{\chi^{\ast}-\chi(z)}{\chi^{\ast}},
\end{equation}
where $\Omega_{m,0}$ is the matter density in units of the critical density at present time, $H(z)$ is the Hubble parameter at redshift $z$ and $H_0$ its value at the present epoch. Once a cosmological model is provided, together with $H(z)$ and $\chi(z)$ relations, then the converge in Equation~\ref{eq:kappa} is determined. \\
On the other hand, radio surveys provide us with the spatial distribution of detected sources but not with that of the underlying matter. Therefore, to link their density contrast in the sky to that of the matter, one has to specify their redshift distribution $N(z)$, that accounts for the survey selection function, and also the bias relation between their spatial distribution and the underlying mass density field.
In this work we assume a linear, scale free, deterministic bias that can be expressed as a function of redshift only: $b(z)$.
The projected density contrast of the radio sources can then be written as
\begin{equation}
\delta_g(\hat{n}) = \int_0^{\infty} b(z) N(z)  \delta_m(\chi(z)\hat{n},z) dz.
\end{equation}

The angular auto power spectrum of the radio sources is obtained by first expanding $\delta_g(\hat{n})$ in spherical harmonics:
\begin{equation}
\delta_g(\hat{n})=\sum_{\ell} \sum_m a_{\ell m}^{g} Y_{\ell m}(\hat{n})\\
\end{equation}
and then by taking the ensemble average of the expansion coefficients, $C^{gg}_{\ell}=\langle a^g_{\ell m} a^{g\,\ast}_{\ell m}\rangle$, so that:
\begin{equation}\label{eq:auto_spectrum_th}
C^{gg}_{\ell}=\frac{2}{\pi} \int_0^{\infty} dz [W^{g}(z)]^2 \int_0^{\infty}dk k^2 P(k,z)j^2_{\ell}[k\chi(z)],
\end{equation}
\begin{figure*}
\begin{multicols}{2}
    \includegraphics[width=\linewidth]{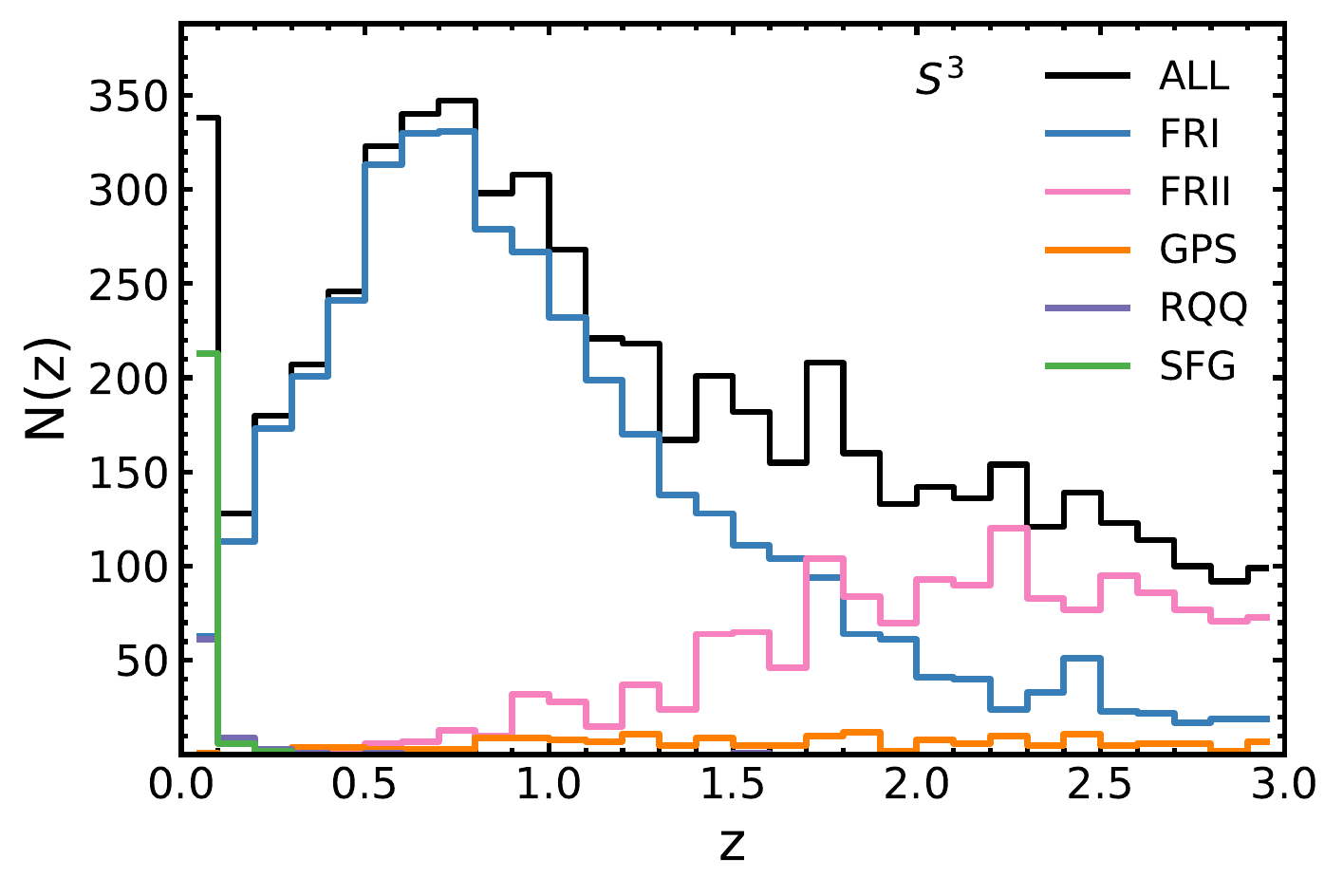}\par
    \includegraphics[width=\linewidth]{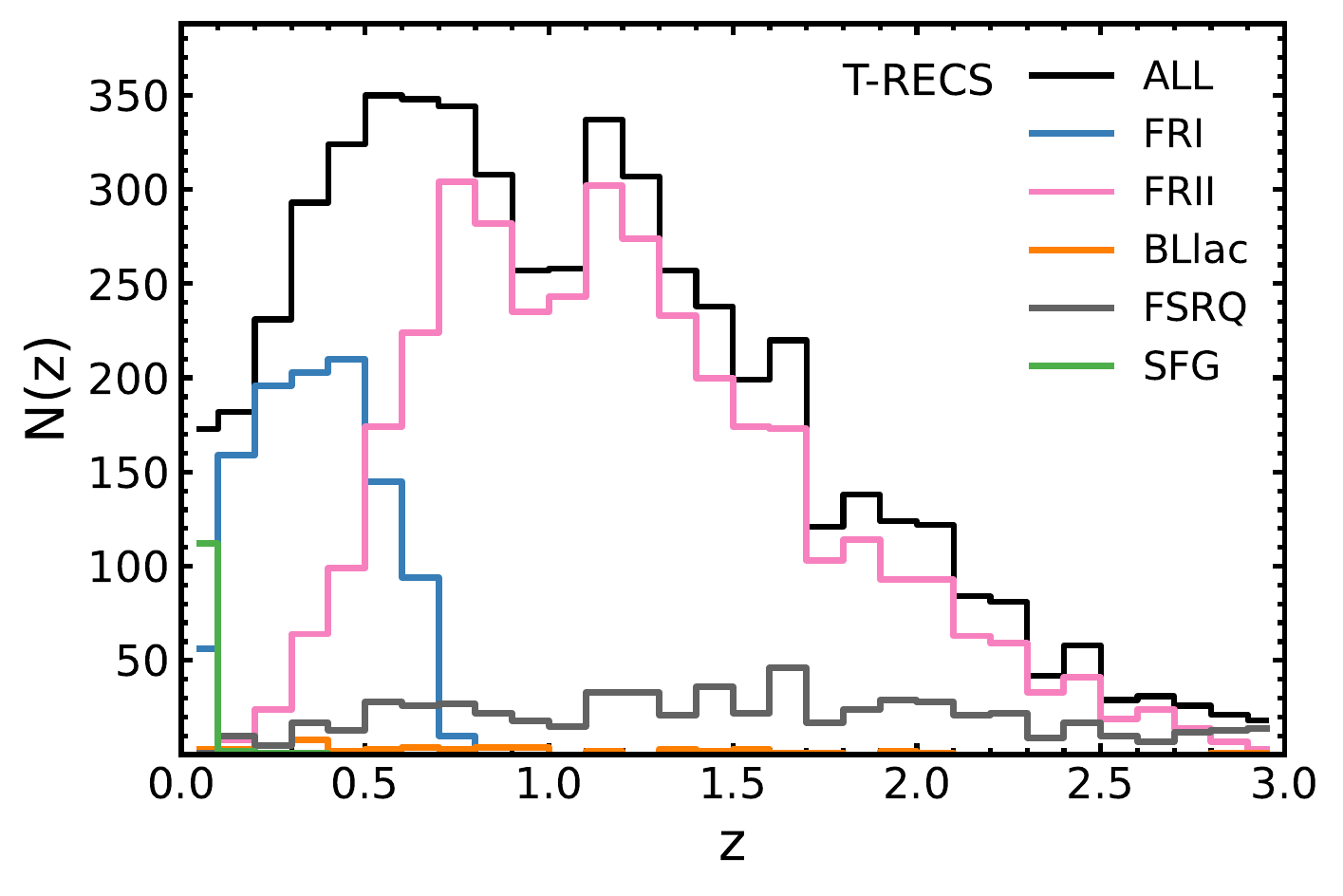}\par 
    \end{multicols}
    \vspace{-1cm}
\begin{multicols}{2}
    \includegraphics[width=\linewidth]{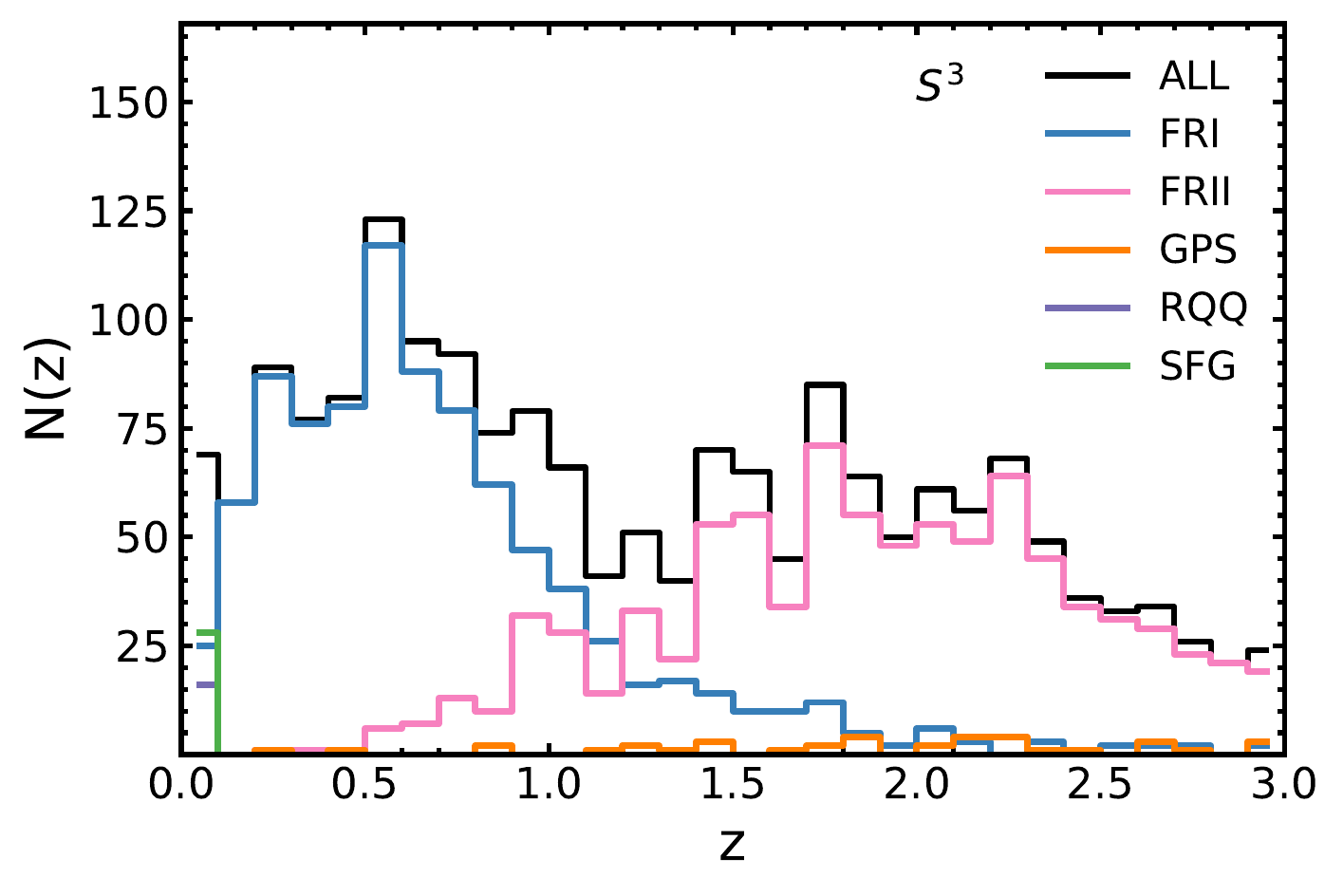}\par
    \includegraphics[width=\linewidth]{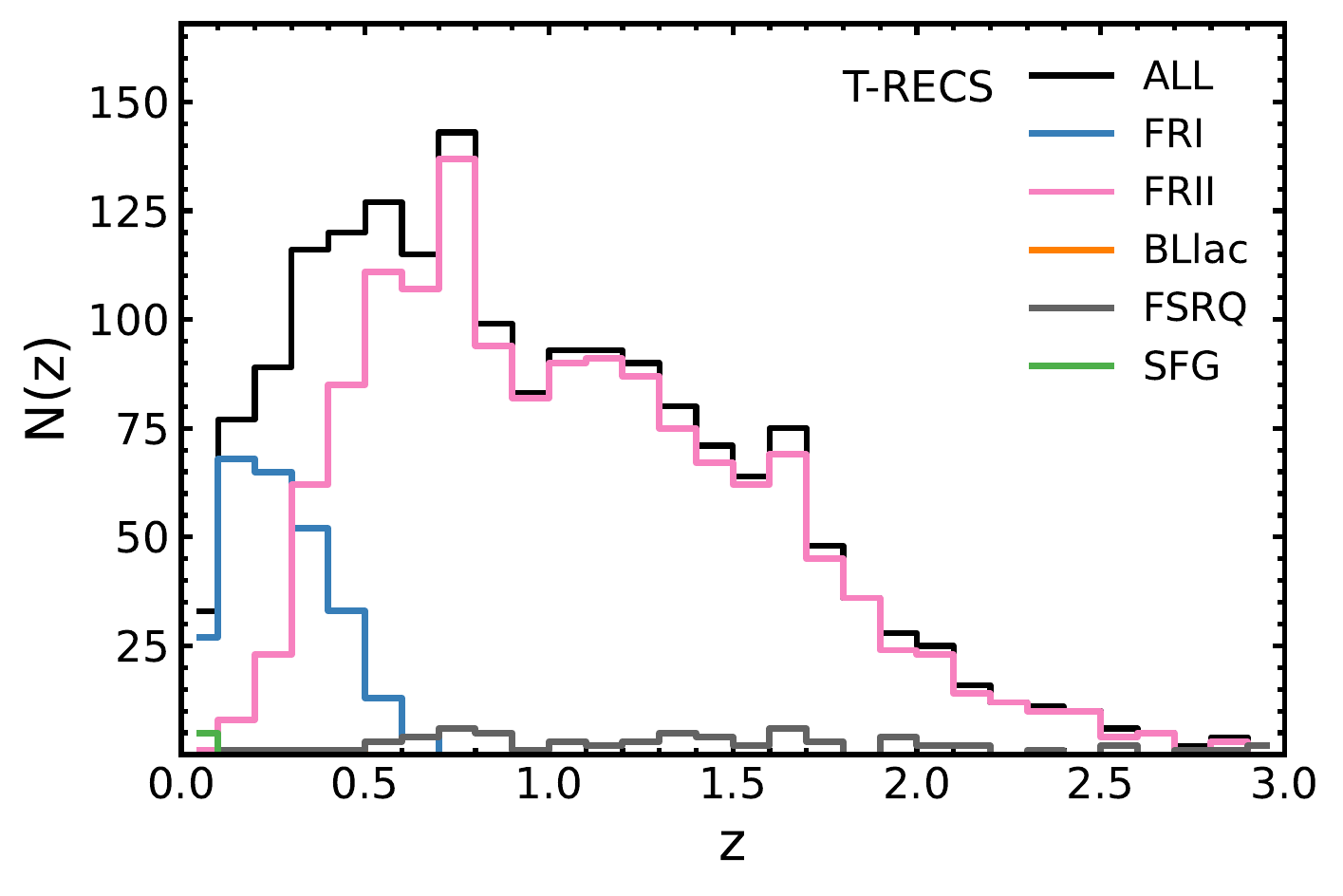}\par
\end{multicols}
\caption{Redshift distribution models $N(z)$ for the NVSS (top panels) and TGSS catalogs (bottom panels) shown in the range $z=[0,3]$. Predictions from the $S^3$ and T-RECS simulators are shown in the plots on the left and on the right, respectively. The thick black histogram in each panel shows the distributions of all sources in the catalog. Histograms with different colors indicate the distribution of each object type identified by the labels.}
\label{fig:nvss_tgss_dndz}
\end{figure*}
where $P(k,z)$ is the three-dimensional power spectrum of the mass density fluctuations at redshift $z$, $j_{\ell}$ are the Bessel functions and $W^{g}(z)$ is the window function of the radio sources which, in our case, is:
\begin{equation}
\label{eq:gal_wind}
W^g(z) = \frac{b(z)N(z)}{\int{dz'{N}(z')}}.
\end{equation}
The CMB lensing convergence - galaxy cross-angular power spectrum is similarly defined as
\begin{equation}
\label{eq:cross_spectrum_th}
\begin{split}
C^{\kappa g}_{\ell} &= \frac{2}{\pi}\int^{\infty}_0 dz   W^g(z)\int^{\infty}_0 dz' W^{\kappa}(z')\times\\
&\int_0^{\infty} dk k^2 P(k,z,z')j_{\ell}[k\chi(z)]j_{\ell}[k\chi(z')],
\end{split}
\end{equation}
where $W^{\kappa}(z)$ and $W^g(z)$ are the window functions defined respectively in Equation~\ref{eq:len_win} and Equation~\ref{eq:gal_wind}. Note that, in the above expression, $P(k,z,z')$ is the matter power spectrum. 

In light of the fact that the relevant multipoles used in our analysis are $\ell \geq 11$ (see Section~\ref{sec:cross_spect} for more details), the theoretical angular cross power spectrum is computed under the so called Limber approximation (\citealt{Limber_1953}). 

Knowing that, the comoving distance $\chi(z)$ is related to the redshift as: $\chi(z) = \int^{z}_0 dz' c/H(z')$, we can
replace the integrals in Equation~\ref{eq:cross_spectrum_th} with a simpler formula:
\begin{equation}
\label{eq:limber_approx}
    C^{\kappa g} = \int^{\infty}_0 \frac{dz}{c}\frac{H(z)}{\chi^2(z)}W^{\kappa}(z)W^{g}(z)P(k=\ell/\chi(z), z).
\end{equation}
The total effect of the Limber approximation is to boost the amplitude of the angular power spectrum at small $\ell$ values. \\
We compute the power spectrum $P(k=\ell/\chi(z),z)$ using the \mintinline{Python}{CAMB}\footnote{\url{https://camb.info/}} code and model nonlinear corrections using the \mintinline{Python}{HALOFIT} approximation (\citealt{Smith2003}).
More specifically, we consider three of the most popular \mintinline{Python}{HALOFIT} versions present in the literature, namely the ones proposed by \cite{Smith2003}, \cite{Takahashi2012} and \cite{Mead2016}. Differences among these approximations are significant only in the highly nonlinear regime, i.e. at large $\ell$ values. After having checked that all the three options provide very similar predictions in the range of multipoles most relevant for our analysis, we decided to rely on the \cite{Mead2016} approximation.
Nonlinear corrections are consistently included also when calculating the effects of CMB lensing, and, also in this case, they have a significant impact only for high values of multipoles.

Note that the integral giving the cross angular power spectrum in Equation~\ref{eq:limber_approx} depends on the product of the two window functions. The radio source window function extends to high redshifts and considerably overlaps with the CMB lensing convergence one, which spans a wide range of redshifts with a broad peak at $z \sim 1$. Because of the significant overlap, we do expect to detect a nonzero cross-correlation signal between the two tracers. The cross-spectrum $C^{\kappa g}_{\ell}$ depends on the bias relation $b(z)$ and on the redshift distribution, $N(z)$, of the radio source. Neither of these quantities are well constrained by observations and they need to be modeled separately, as we describe below.

For the redshift distributions of radio sources we rely on two models. The first one is obtained from the SKA Simulated Skies database ($S^3$ hereafter, \citealt{Wilman_2008}). The $N(z)$ of the $S^3$ simulator is designed to match the luminosity function of different types of radio sources observed  at different redshifts. The total $N(z)$ model is shown in the two left panels of Figure~\ref{fig:nvss_tgss_dndz} with a thick black line histogram. Being the $N(z)$ of a composite sample, the model also provides the $N(z)$ of the various types of sources in the catalog, namely: Fanaroff-Riley class I sources (FRI, blue), Fanaroff-Riley class II sources (FRII, pink), GHz-peaked radio sources (GPSs, orange), radio quiet quasars (RQQs, purple) and star forming galaxies (SFGs, green). Top and bottom panels refer to the NVSS and TGSS samples, respectively. The second $N(z)$ model is obtained from the Tiered Radio Extragalactic Continuum Simulation (T-RECS) of \cite{Bonaldi_2018} where radio sources are assigned to the dark matter halos in the light cone and constrained to match the luminosity functions and the clustering properties of Active Galactic Nulcei (AGN) and SFGs. The T-RECS $N(z)$ models of the NVSS (top) and TGSS (bottom) are shown in the right panels of Figure~\ref{fig:nvss_tgss_dndz}. The sources in the T-RECS simulation have a slightly different classification which includes BL Lac objects (BLlac, orange histogram) and Flat-Spectrum Radio Quasars (FSRQ, grey) in addition to SFGs, FRIs, and FRIIs.
It is worth to point out that the redshift distributions and the source compositions predicted by the two models are significantly different. $S^3$ predicts a higher fraction of SFGs at low redshifts and a more extended high-redshift tail than T-RECS. Note that in our analysis, we consider the $N(z)$ distributions up to $z=3$, since the luminosity functions of the astrophysical sources are quite uncertain beyond that redshift (i.e. \citealt{Allison_2015}).

Both NVSS and TGSS are composed by different types of sources. The resulting effective bias of the full sample can be defined as:
\begin{equation}
\label{eq:bias_eff}
b(z) \equiv \frac{\sum_i N_i(z)b_i(z)}{\sum_i N_i(z)},
\end{equation}
where $b_i(z)$ and $N_i(z)$ are the galaxy bias and the redshift distribution of each object type.
In this work, we consider six different models for the effective bias $b(z)$, 
divided into two categories based on whether they are taken from the literature or estimated from the data.

\begin{figure}
\centering
\includegraphics[width = \hsize]{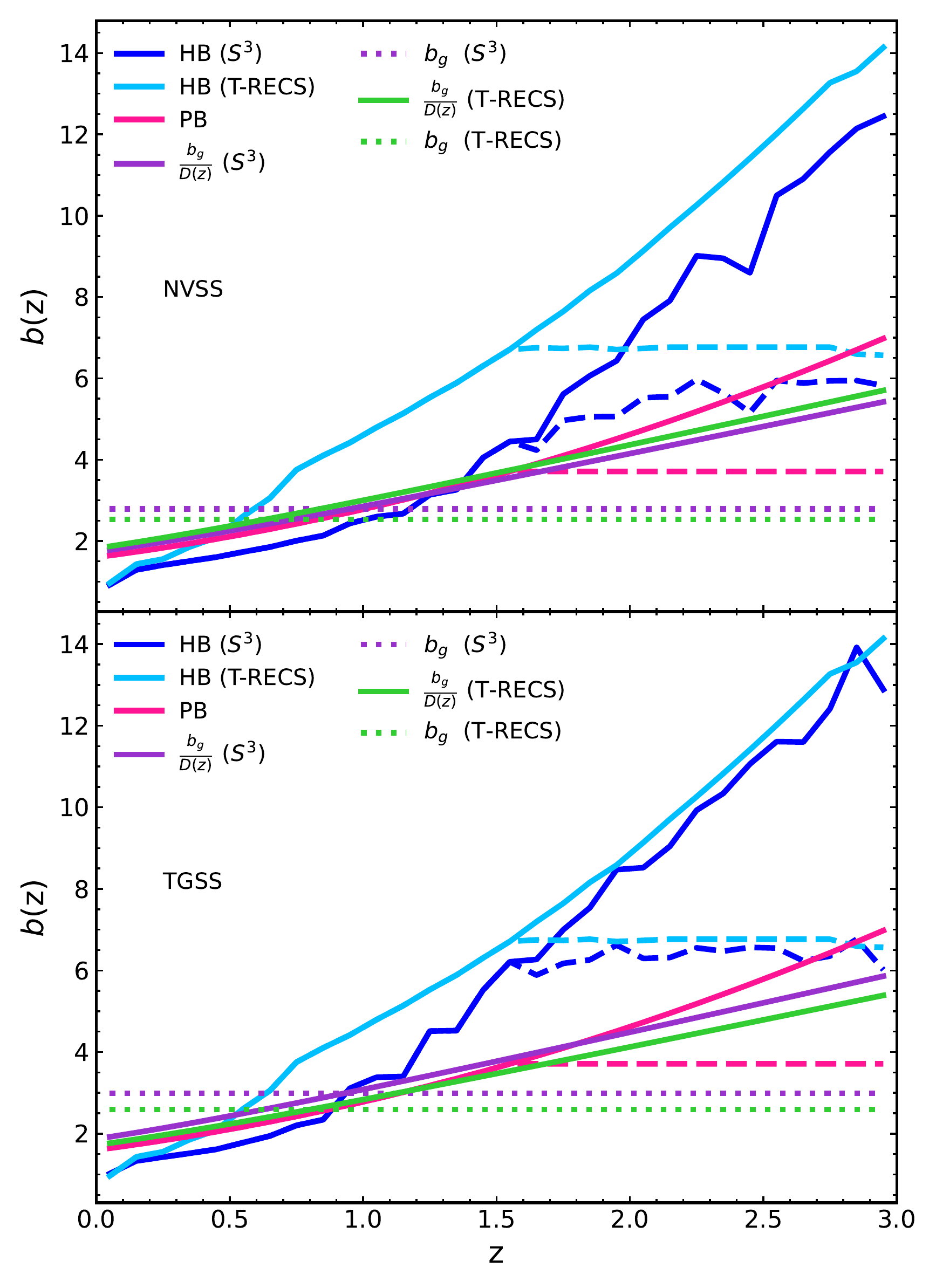}
\caption{Bias models used in this work as a function of redshift in the range $z = [0,3]$ for the NVSS (top panel) and TGSS (bottom panel) cases. Curves with different colors and line styles are used for the different combinations of $N(z)$ and galaxy bias models, as indicated in the labels. 
The values of the free parameter $b_g$ are listed in Table~\ref{tab:tgss_kappa_bias_bf} and in Table~\ref{tab:nvss_kappa_bias_bf} for TGSS and NVSS respectively. We also consider "truncated" versions of the HB and PB models in which the value of the bias for each population of sources is kept constant for $z \geq 1.5$ (dashed lines).}
\label{fig:nvss_tgss_s3_trecs_bz}
\end{figure}

The first category of bias prescriptions is based on the halo model (HM hereafter, \citealt{COORAY_2002}) and they have been proposed in previous studies to describe the auto-correlation function of the radio sources. 
The "Halo Bias" (HB) model was presented by \cite{ferramacho2014} and used by \cite{dolfi_2019} in the NVSS and TGSS auto-correlation studies.
It uses the HM prescription to assign different types of radio sources to halos of different masses and, accordingly, with different biases $b_i(z)$. As a consequence, looking at Equation~\ref{eq:bias_eff} the HB model depends on the assumed $N(z)$, either $S^3$ or T-RECS. The two corresponding HB models are shown in Figure~\ref{fig:nvss_tgss_s3_trecs_bz} with the deep and light blue curves for the NVSS (top panel) and TGSS (bottom panel) case. Both models predict that the bias steadily increases with the redshift, which is rather nonphysical. 
Therefore, we also consider a "Truncated Halo Bias" (THB) model in which we set the bias of each source population to remain constant for $z\geq 1.5$. This is shown as a dashed curve. The "Parametric Bias" (PB) model has been proposed by \cite{Nusser_2015}
to describe the angular power spectrum of NVSS sources. Following \cite{dolfi_2019}, we also use the same PB model in combination with both T-RECS and $S^3$ distributions, for the NVSS and TGSS catalogs. It is represented by the magenta curves in the panels of Figure~\ref{fig:nvss_tgss_s3_trecs_bz}. As for the HB prescription, we also consider a truncated version (TPB) with constant bias for $z\geq 1.5$.

The second category of bias models are simpler and are characterized by parameters that are free to best fit the data. \textit{Model 1} assumes that the bias of radio sources evolves with the inverse of the linear growth factor: $b(z) = b_g/D(z)$, where $ b_g$ is the free parameter of the model and $D(z)$ has been computed using the \mintinline{Python}{Colossus}\footnote{\url{https://bdiemer.bitbucket.io/colossus/}} toolkit assuming the reference Planck $\Lambda$CDM cosmology \citep{planck_params}. In Figure \ref{fig:nvss_tgss_s3_trecs_bz} the model is shown for the $S^3$ and T-RECS $N(z)$ cases with the solid purple and light green curves, respectively.
Bias \textit{model 2}, instead, assumes a constant bias $b(z) = b_g$ and it is shown with dotted curves. The $b_g$ values used in the figure are those that best fit the data (see Section~\ref{sec:bias_fit}, Table~\ref{tab:tgss_kappa_bias_bf} and Table~\ref{tab:nvss_kappa_bias_bf}).

To summarize, the ingredients used to model the auto- and the cross-angular spectra are: the reference $\Lambda$CDM background cosmology \cite{planck_params}, a \mintinline{Python}{HALOFIT}-based matter power spectrum $P(k,z)$ model
generated with \mintinline{Python}{CAMB}, a  model for the redshift distribution of radio sources, $N(z)$, and for their bias, $b(z)$. 
We stress that, these two quantities take into account the composite nature of the sample by considering the $b_i(z)$ and $N_i(z)$ functions of each radio type (see Equation~\ref{eq:bias_eff}). 
We used these quantities to model the angular spectra in Equations~\ref{eq:auto_spectrum_th} and \ref{eq:cross_spectrum_th} since the \mintinline{Python}{CAMB} version used in this work does not offer the possibility to include the contribution of each population separately.

For reasons that will be justified in Section~\ref{sec:cross_spect}, in our analysis we only consider multipoles $\ell \geq 11$. 
As a result the effect of peculiar motions, which boosts the clustering amplitude on large angular scales, is negligible, as we have verified by including this effect in one of our model spectra (the one that adopts the TPB+$S^3$ model combination).
In this case, that we regard as representative, the boosting effect at $\ell \simeq 10$ is $0.7\%$.
A second benefit is the possibility of using the Limber approximation. Its impact has been evaluated for the same TPB+$S^3$ model. The difference between angular spectra evaluated using the Limber approximation and through a complete 2D integral is less than $0.2\%$ at $\ell \simeq 10$, decreasing thereafter.
Finally, we also include the effect of the lensing magnification bias which systematically modifies the number of objects above the flux threshold of the catalog. The magnitude of the effect depends on the effective slope $\tilde{\alpha}$ of the luminosity function of radio sources in the faint end. It can be estimated by combining the slopes of the luminosity functions of the individual objects' type, as (\citealt{dolfi_2019}):
\begin{equation}
\label{eq:magnification_bias}
\tilde{\alpha}=\frac{\sum_i \sum_j \alpha(i,j)N_i(z_j)}{\sum_i \sum_j N_i(z_j)}\simeq 0.30,
\end{equation}
where $j$ runs over the redshift bins, $i$ identifies the object type, $N_i(z)$ its redshift distribution and $\alpha(i,j)$ is the faint-end slope of each luminosity function type $i$ at the redshift $j$. The sensitivity of our results to lensing magnification bias will be assessed in 
Section~\ref{sec:magbias}.

\section{Estimated angular power spectra}\label{sec:estimators}
We measure the auto- and cross-correlation of the CMB lensing convergence $\kappa$ and galaxy counts $g$ in harmonic space, using the angular {\it pseudo}-power spectrum formalism.

Let us expand a generic field $X$ defined on a pixelized 2D map over a fraction of the sky in spherical harmonics:
\begin{equation}\label{eq:pseudo_alm}
\tilde{a}^X_{\ell m} = \sum_{i=1}^{N_{pix}} X_i\, w^X_i\, Y^*_{\ell m}(\theta_i) ,\\\nonumber
\end{equation}
where the sum runs over the pixels. The weight function $w^X_i$ quantifies the effect of the mask described in Section~\ref{sec:data}. It is set to zero in unobserved sky areas or in pixels in which the signal-to-noise is below some threshold.  

Under the assumption of statistical isotropy, the {\it pseudo} angular power spectrum of two fields $X,Y=\{\kappa, g\}$ can be estimated from the measured harmonic coefficients as: 
\begin{equation}\label{eq:pcl_def}
\tilde{C}^{XY}_{\ell}  = \frac{1}{2\ell+1}\sum_{m=-\ell}^{+\ell} \tilde{a}^X_{\ell m}\;\tilde{a}^{Y*}_{\ell m}\,.
\end{equation}
When $X=Y$ we get the {\it auto} angular power spectrum, while if they differ the expression provides the {\it cross} angular power spectrum. 
The actual power spectrum is derived from the measured {\it pseudo}-spectrum as \citep{MASTER, MASTER2}:
\begin{equation}\label{eq:cb_est_gen}
\hat{C}^{XY}_\ell=\sum_{\ell'}M^{-1}_{\ell \ell'}\tilde{C}^{XY}_{\ell'}- {N}^{XY}_{\ell}.
\end{equation}
In the above equation, ${N}^{XY}_{\ell}$ is an estimate of the noise angular power spectrum, which 
is non-zero only when $X=Y$. 
The matrix $M^{XY}_{\ell \ell'}$  accounts for the power-loss and  mode-mode coupling 
due to incomplete sky coverage and survey footprint.
It is defined as:
\begin{align}\label{eq:kern_def}
  M^{XY}_{\ell \ell'} & =  \frac{(2 \ell' + 1)}{ 4 \pi}\sum_{\ell'' } (2 \ell'' + 1)\;\tilde W^{XY}_{\ell''}\;
{\left ( \begin{array}{ccc}
        \ell & \ell' & \ell''  \\
        0  & 0 & 0
       \end{array} \right )^2}. \\\nonumber
\end{align}
Here $\tilde W^{XY}_\ell$ is the cross-power spectrum of the masks which is defined as:
\begin{equation}\label{eq:wf_ps}
 \tilde W^{XY}_\ell = \frac{1}{2 \ell + 1} \sum_{m=-\ell}^{\ell} \tilde w^{X}_{\ell m}\; \tilde w^{Y*}_{\ell m},
\end{equation} 
where $\tilde w^{X}_{\ell m}$ and $\tilde w^{Y}_{\ell m}$ are the spherical harmonic coefficients of the masks of the analysed fields. 

Under the assumption that $\kappa$ and $g$ are both random variables, we can represent the joint covariance matrix of the angular auto $gg$ and cross $\kappa g$ spectra as:
\begin{equation}
\label{eq:cov_joint}
\mathrm{Cov}^{joint} =
\begin{bmatrix}
\mathrm{Cov^{\kappa g, \kappa g}} & (\mathrm{Cov^{\kappa g, gg}})^T \cr
\mathrm{Cov^{\kappa g, gg}} & \mathrm{Cov^{gg, gg}}
\end{bmatrix},
\end{equation}
in which each block can be written as
\begin{align}\label{eq:diag_cov} \nonumber
{\rm Cov}^{Xg,\, X'g}_{\ell \ell'} = &\frac{\delta_{\ell \ell'}}{(2\ell+1)\,f^{Xg,\,X'g}_{\rm sky}}
\left[\left({C}^{XX'}_\ell+{N}^{XX'}_\ell\right)\left({C}^{gg}_{\ell'}+{N}^{gg}_{\ell'}\right) \right.\\
& + \left.\left({C}^{Xg}_\ell+{N}^{Xg}_\ell\right)\left({C}^{gX'}_{\ell'}+{N}^{gX'}_{\ell'}\right)\right]. \
\end{align}
In the above Equation, $X$ and $X'$ can be both $\kappa$ and $g$, ${C}_\ell$s are the theoretical angular power spectra corresponding to the fiducial model, while the sky fractions are defined as $f^{Xg,\,X'g}_{\rm sky}= \sqrt{f^{Xg}_{\rm sky}\cdot f^{X'g}_{\rm sky}}$ and the different $f^{Xg}_{\rm sky}$ values are given in Section \ref{sec:data}

%
%

In our analysis, we bin the estimated spectra $\hat{C}^{XY}_\ell$ within equally spaced linear bins $\Delta \ell$. Consequently, the elements of the covariance matrix will also consist of $\Delta \ell$-binned quantities. 
The main reason for binning the spectra is to reduce the multipole covariance induced by the mask hence decreasing the amplitude of the off-diagonal elements. 
In our main analysis, we set the bin size to $\Delta \ell=30$. This choice guarantees that the covariance matrix is well approximated by a diagonal Gaussian model, as we show in Appendix~\ref{sec:cov_appendix}. In the same Appendix, using Monte Carlo simulations, we also verify that our pipeline for power spectrum estimation is unbiased, namely it does not introduce spurious correlations and it is able to recover a known input power spectrum. 

\subsection{Angular auto-power spectrum of radio sources}
\label{sec:auto_spect}
To measure the angular power spectrum of the radio sources, $\hat{C}^{gg}_{\ell}$, we use the estimator defined in Equation~\ref{eq:cb_est_gen} with  $X=Y=g$ which also corrects for a noise term. For the auto-spectrum it is given by the sum of the Poisson noise $1/\overline{N}$ ($\overline{N}$ being the mean number of radio sources per pixel) and the spurious contribution due to multiple components of a single source counted as individual sources. 
These noise terms were estimated by \citealt{dolfi_2019} for the TGSS and NVSS samples used in their analysis.

We re-assess these corrections by requiring that the auto-spectrum should be consistent with zero at high $\ell$ where the noise term is expected to be dominant. We find that the original \citealt{dolfi_2019} correction is adequate for the TGSS case and it is $N^{gg} = 8.86 \times 10^{-5}$. On the other hand, for the NVSS case it provides an over correction and, as a consequence, the angular auto-spectrum takes negative values at high multipoles.
For this reason, we estimate the correction from the data by treating the noise term $N^{gg}$ as a free parameter in the best fitting procedure described in Appendix \ref{sec:shot_noise_estimation}. With this method, the best fitting noise terms listed in Table~\ref{tab:shot_values} depend (only mildly) on the theoretical model adopted for the angular power spectrum.

 

\begin{figure}
    \centering
    \includegraphics[width=\hsize]{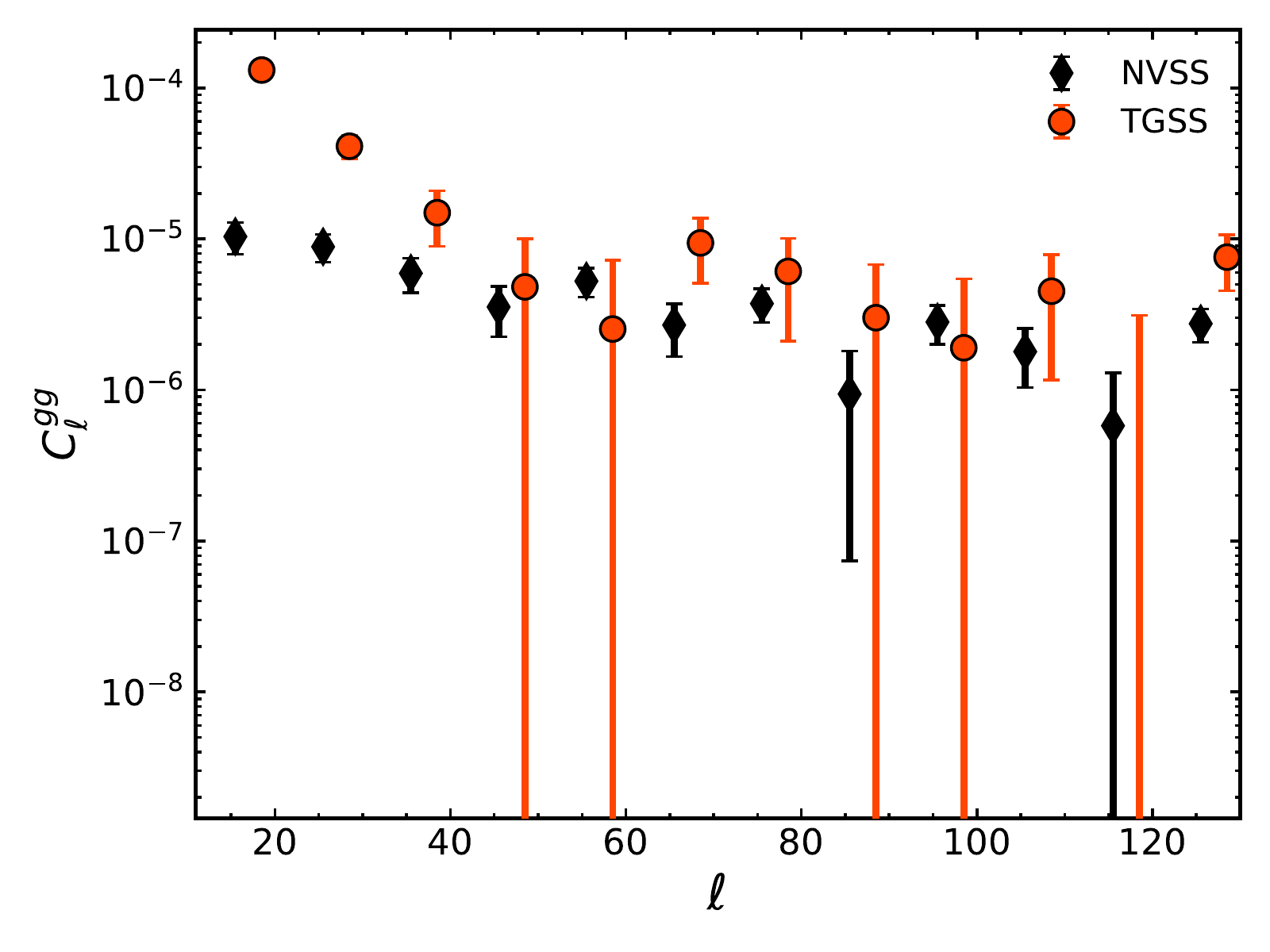}
    \caption{Binned angular power spectra of NVSS (black diamonds) and TGSS (red dots) sources in the range $\ell=[11, 130]$. The size of the bins is $\Delta_{\ell}=10$. Error bars are 1-$\sigma$ Gaussian uncertainties from  Equation~\ref{eq:diag_cov}. Power spectra are corrected for both shot noise and multiple components contributions (see text for details).
    }
\label{fig:auto_spect}
\end{figure}

The binned angular power spectra measured in the range $\ell=[11,130]$ for the reference NVSS and TGSS samples are shown in Figure \ref{fig:auto_spect} with black diamonds and red dots, respectively. The minimum value $\ell_{min}=11$ allows us to exclude some problematic multipoles which are affected by a varying flux sensitivity that depends on the Galactic latitude of the observed NVSS sources (\citealt{Smith_2007}). The choice of $\ell_{max} = 130$ is due to the fact that, for greater multipoles, the auto-spectra are dominated by shot noise. We also use the same $\ell$-cuts for TGSS. Errorbars in the plot represent the 1-$\sigma$ uncertainties computed with Equation~\ref{eq:diag_cov}, for $X=X'=g$.
Note that the binning scheme used for this plot is $\Delta_{\ell}=10$ to ease the comparison with the auto-spectra estimated by \cite{dolfi_2019}. This outcome confirms their result and the power excess in the TGSS sample for $\ell \leq 20-30$. 

\subsection{Angular cross-power spectrum of CMB lensing convergence and radio sources}
\label{sec:cross_spect}
The cross power spectrum $\hat{C}_{\ell}^{\kappa g}$ is estimated using Equation~\ref{eq:cb_est_gen} with $X=\kappa $ and $Y=g$. We estimate the cross-spectrum in the multipole range  $\ell=[11,310]$ and in 
bins $\Delta_{\ell}=30$.  
The $\ell$-range is wider with respect to the auto-spectrum case. Its upper 
value $\ell_{max} = 310$ is set to minimize the impact of nonlinear effect that would break the Gaussian hypothesis adopted to model the covariance matrix. To estimate $\ell_{max}$, we compare the theoretical angular cross-spectrum obtained assuming a linear matter power spectrum 
with that obtained using the \mintinline{Python}{HALOFIT} approximation and search for the $\ell$ value at which the difference between linear and nonlinear predictions reach $10\%$.
As we considered three different \mintinline{Python}{HALOFIT} models, we obtained three values for $\ell_{max} = 310, \, 325, \, 350$ and opted for the most conservative one. We also notice that the gain in signal-to-noise is quite low above this multipole.

\begin{figure}
    \centering
    \includegraphics[width=\hsize]{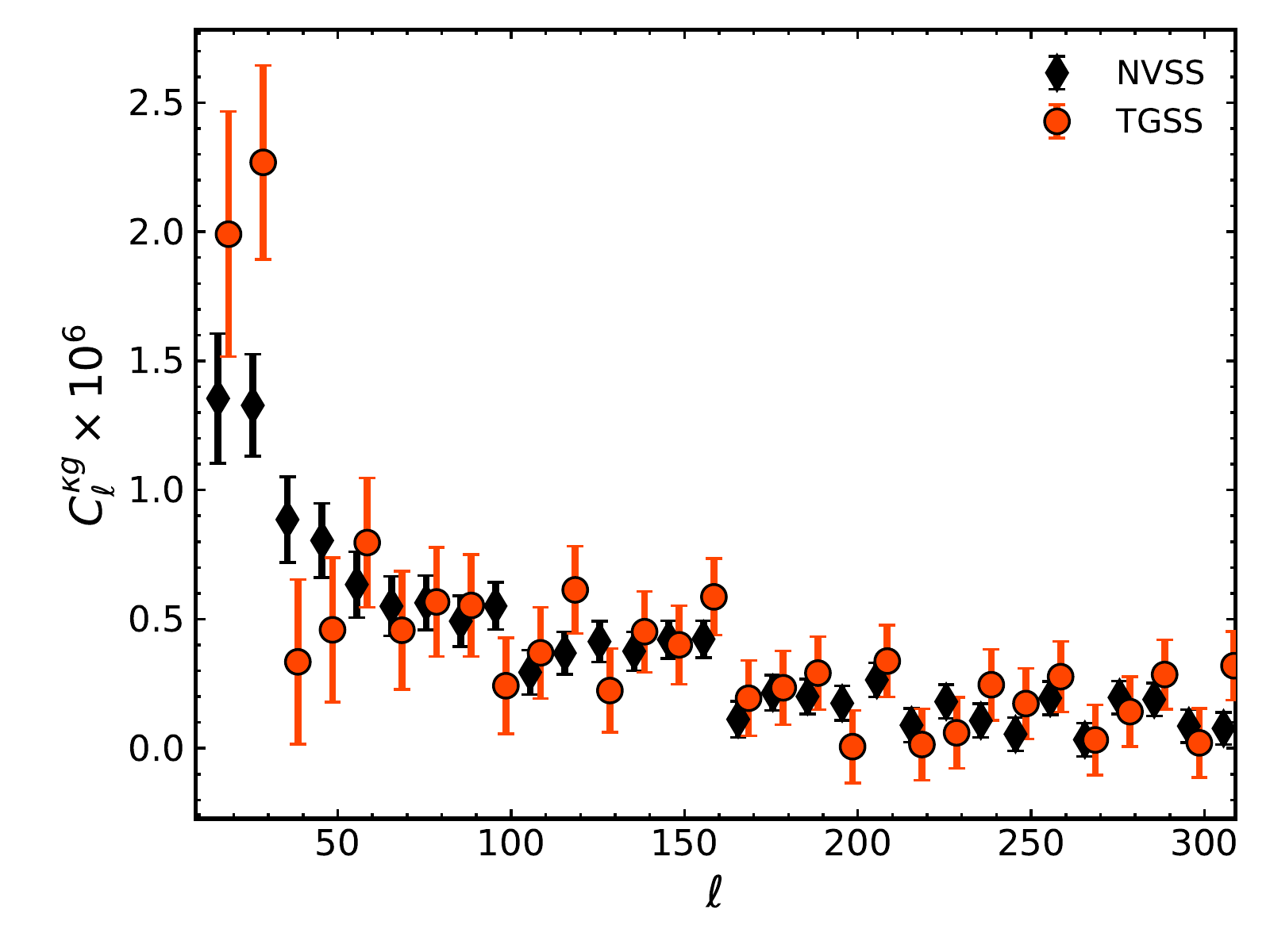}
    \caption{Cross angular power spectra of the NVSS catalog (black diamonds) and TGSS catalog (red dots) with CMB lensing convergence in the range $\ell=[11,310]$ measured in bins $\Delta_{\ell} = 10$. Errorbars are 1-$\sigma$ Gaussian uncertainties from Equation~\ref{eq:diag_cov}.}
    \label{fig:tgss_nvss_cross}
\end{figure}

The estimated CMB lensing convergence-radio source cross-spectrum is shown in Figure~\ref{fig:tgss_nvss_cross} for the NVSS (black diamonds) and TGSS (red dots) cases. Errorbars represent 1-$\sigma$ uncertainties obtained with Equation~\ref{eq:diag_cov} where $X=X'=\kappa$. 
Unlike for the auto-spectrum case, we now find a good match between the two samples down to the smallest multipole explored, i.e. we do not find evidence of a power excess in the spectrum of TGSS sources anymore.
This result strongly suggests that the power excess in the TGSS auto-spectrum is not genuine but likely originates from uncorrected systematic effects in the observations. Hence, it also corroborates the suggestion made by \cite{Tiwari_2019} that the excess power in the TGSS catalog probably originates from systematic effects in the flux calibration. Moreover, this result clearly illustrates the effectiveness of the cross-correlation as a tool to identify and reduce the impact of any systematic effect that does not correlate with the LSS.
 
Note that for both radio catalogs, we report a high-significance detection of the cross-correlation with Planck CMB lensing. To assess the significance of this detection, we compare our results with the null hypothesis, namely the probability that no correlation is found between the CMB lensing convergence field and the radio galaxies distribution. We quantify this probability by estimating $\chi^2_{null}$ :
\begin{equation}
    \chi^2_{null} = \sum_{\ell, \ell'}\hat{C}^{\kappa g}_{\ell}(\mathrm{Cov^{\kappa g, \kappa g}_{\ell \ell'}})^{-1}\hat{C}^{\kappa g}_{\ell'},
\end{equation}
where $\hat{C}^{\kappa g}_{\ell}$ is the estimated power spectrum and $\mathrm{Cov^{\kappa g, \kappa g}_{\ell \ell'}}$ is the covariance matrix associated with the cross power spectrum only. We obtain the significance in unit of sigma as the square root of the difference between the null hypothesis and the $\chi^2$ of the best fit model, namely: $\sqrt{\chi^2_{null}-\chi^2}$. The best-fit theoretical model is the one obtained considering TPB + T-RECS as it is reported in Table~\ref{tab:chi2_tgss} and Table~\ref{tab:chi2_cross_nvss} respectively for TGSS and NVSS (see Section~\ref{sec:modelvsdata} for more detailed analysis and notation). We found a detection significance of $22\sigma$ for the NVSS catalog, compatible with \citep{planck_lens2013}, and of $12\sigma$ for TGSS.

\section{Testing radio sources \texorpdfstring{$b(z)$}{b(z)} and \texorpdfstring{$N(z)$}{N(z)} models}
\label{sec:modelvsdata}
We now compare the angular spectra estimated in the previous Section 
with model predictions obtained using different prescriptions for the redshift distribution and for galaxy bias of the radio sources. Some of these prescriptions are taken from the literature, some are firstly proposed here, as it is pointed out in Section~\ref{sec:theory}.
We perform two different analyses. The first one involves cross-spectra only. The second one is a joint analysis that includes both cross- and auto-spectra. In both cases we use a $\chi^2$ statistics to assess the relative goodness of the proposed models:

\begin{equation}
\chi^2 = \sum_{\ell, \ell'} \left(C_{\ell}-\hat{C}_{\ell}\right)
\mathrm{Cov^{-1}_{\ell \ell'}}\left(C_{\ell'}-\hat{C}_{\ell'}\right),
\label{eq:chi2}
\end{equation}
where the sum runs over the multipole bins, $\hat{C}_{\ell}$ is the data vector, $C_{\ell}$ represents the model and $ \mathrm{Cov_{\ell \ell'}}$ is the covariance matrix. For all $\chi^2$ analyses we consider bins $\Delta_{\ell}=30$. To evaluate the consistency of the data vector with the model predictions we provide, along with the reduced $\chi^2$, the probability-to-exceed (PTE) i.e. the probability of obtaining a $\chi^2$ value higher than what we actually measure: ${\rm PTE}=1-P(< \chi^2)$. 

In the first $\chi^2$ analysis, the data vector is the estimated cross-spectrum defined in Section~\ref{sec:cross_spect} $\hat{C}^{\kappa g}_{\ell}$, the model power spectrum is that of Section~\ref{sec:theory} $C^{\kappa g}_{\ell}$ and we use the Gaussian diagonal covariance matrix of Equation~\ref{eq:diag_cov} where $X=X'=\kappa$. For the cross-spectrum we consider 10 equally spaced bins in the range $\ell=[11,310]$. As a result we have a covariance matrix of $10\times10$ elements.

The second analysis includes both the $\kappa g$ cross- and the $g g$ auto-spectra. The latter has been estimated in Section~\ref{sec:auto_spect}. Since we use a $\Delta_{\ell}=30$ binning scheme, it contributes with additional 4 elements to the data vector that now contains 14 elements.
The size of the corresponding covariance matrix is then $14\times14$.
Our results for the two analysis are further discussed in the following Sections.

\subsection{Cross angular power spectrum results}
\label{sec:chi2_cross_spectra}
\begin{figure}
\centering
\subfloat{\includegraphics[width=\hsize]{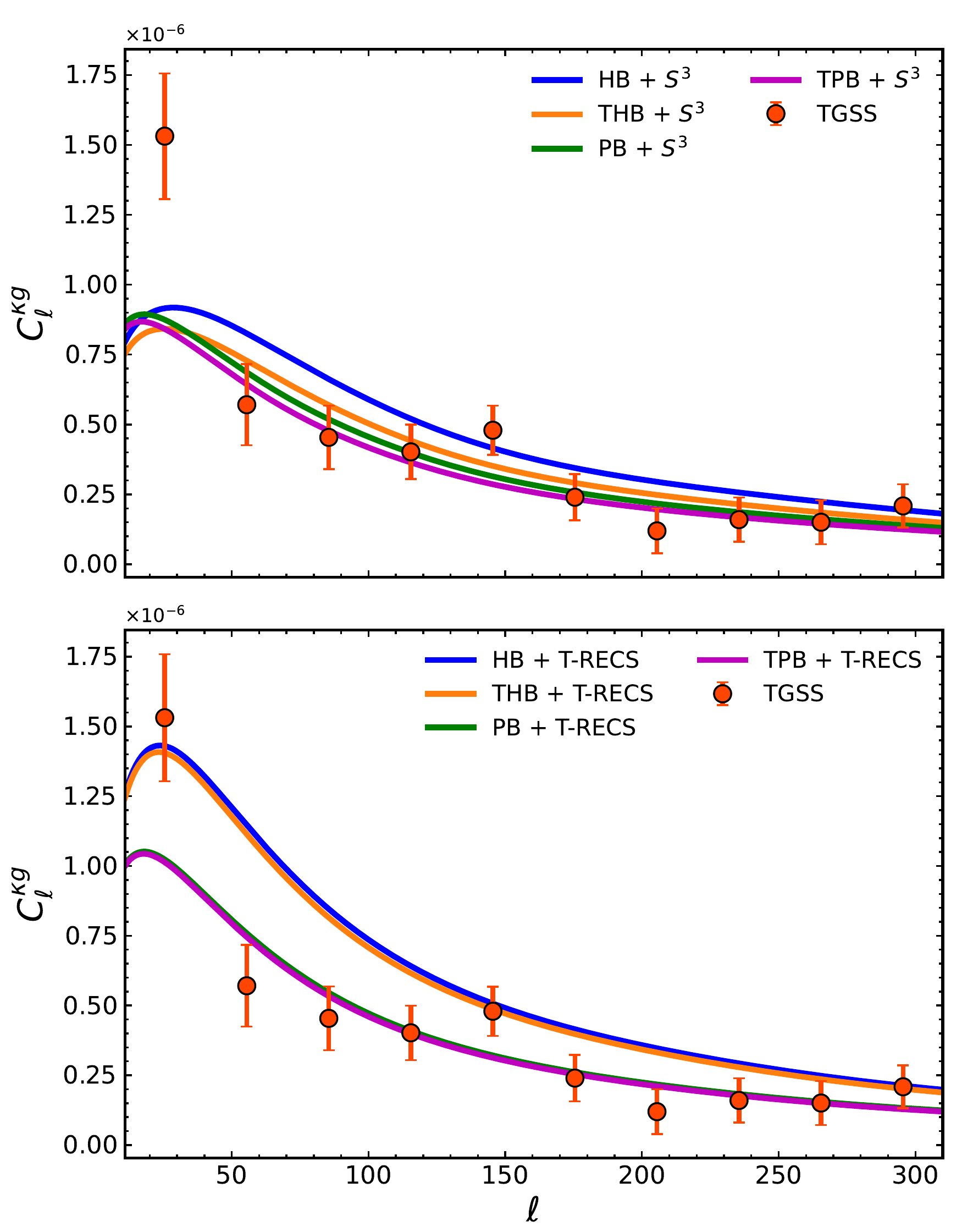}}\hfill
\caption{$\kappa g$ cross-spectrum analysis for the TGSS sample. Errorbars represent 1$\sigma$ Gaussian uncertainties. Curves with different colors: theoretical predictions of different $b(z)-N(z)$ combinations specified by the labels. Top panel: models that use the $S^3$ $N(z)$. 
Bottom panel: models that use the T-RECS $N(z)$ predictions. 
}
\label{fig:tgss_cross}
\end{figure}

In the two panels of Figure~\ref{fig:tgss_cross}, we compare the measured TGSS-CMB lensing convergence cross-spectrum 
with predictions obtained from different $N(z)$ and $b(z)$ models, as summarized by the labels.
We distinguish two sets of models. Those obtained assuming the $N(z)$ simulated by $S^3$ (top panel) and those obtained using the $N(z)$ simulated by T-RECS (bottom). Errorbars represent 1$\sigma$ Gaussian uncertainties. Since the covariance matrix is computed for each model, they are model dependent. However, as we verified, the dependence is weak and we plot those obtained with the TPB bias model for reference.
The results of the $\chi^2$ analysis are reported in
Table \ref{tab:chi2_tgss}, where we list the values of the reduced $\chi^2$ which have been obtained for 10 degrees of freedom (d.o.f.), together with the corresponding PTE values.

\begin{figure}
    \centering
    \includegraphics[width=\hsize]{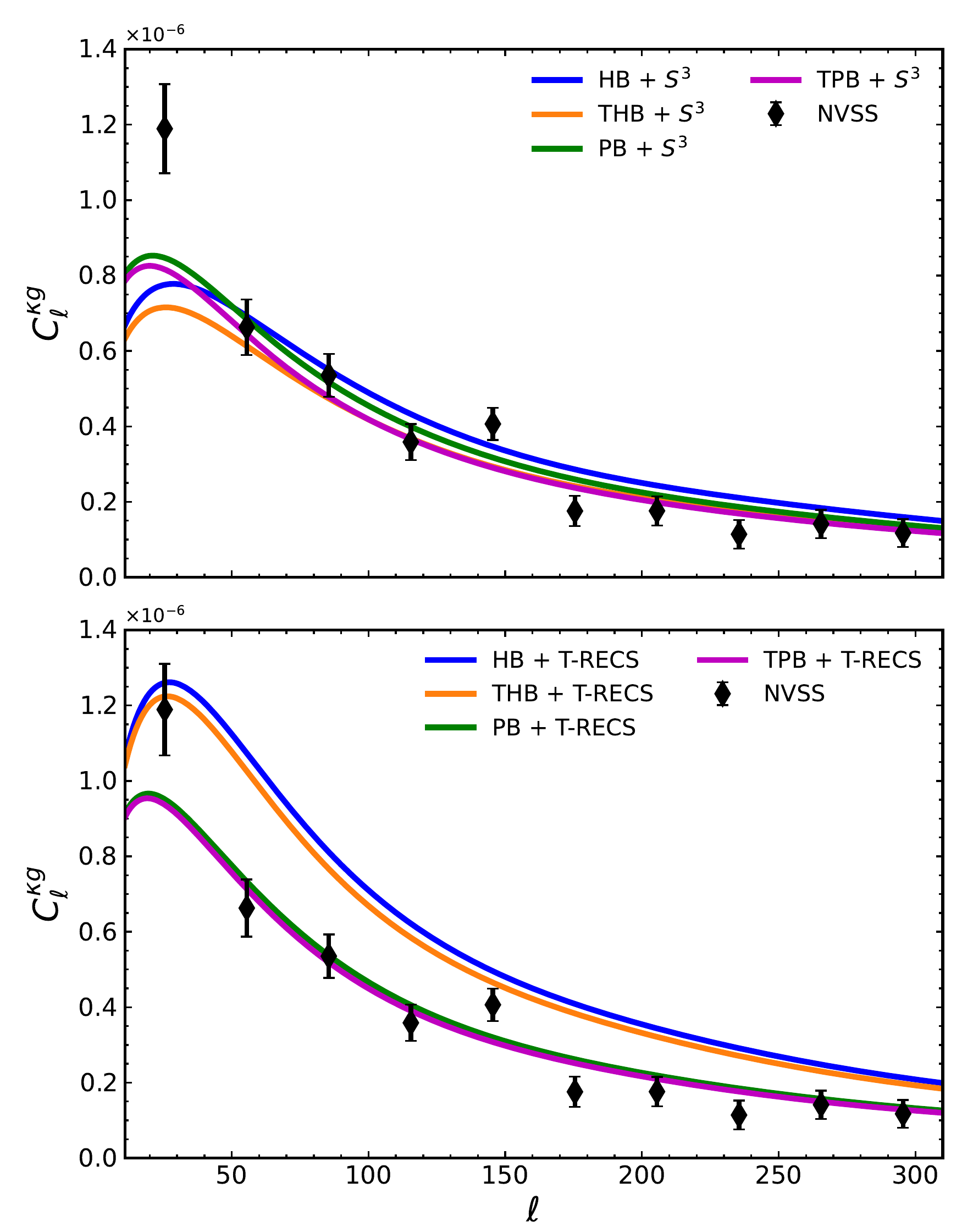}
    \caption{Same as Figure~\ref{fig:tgss_cross} but for the NVSS $\kappa g$
    cross-spectrum analysis. Errorbars represent 1$\sigma$ Gaussian uncertainties.}
    \label{fig:nvss_cross}
\end{figure}

The visual inspection of Figure~\ref{fig:tgss_cross} reveals several interesting features. First, for a fixed bias model, i.e. the PB and TPB cases, the amplitude of the cross spectrum obtained in the T-RECS case is systematically larger than in the $S^3$ case.
This reflects the different redshift distributions and biasing properties of the radio sources in the two $N(z)$ models. According to T-RECS, TGSS is dominated by a population of bright FRII sources with an effective bias that rapidly increases with the redshift.
In the $S^3$ case, the fraction of FRII sources is smaller. Their distribution extends to higher redshifts and at moderate redshift the sample is dominated by a comparatively fainter FRI + SFG population whose bias increases slowly with the redshift. A second remarkable feature is the impossibility, for all $S^3$ models, to match the amplitude of the cross-spectrum in the first bin. A match that is obtained when using the T-RECS $N(z)$ and the HB and THB bias models. Again, this reflects the fact that the population of nearby objects, that most contribute to the amplitude of the spectrum on large angular scales, in the $S^3$ case is dominated by object that are less biased than their T-RECS counterparts. We should also notice that while T-RECS + HB and T-RECS + THB fit the amplitude of the first bin, they fail to match the cross-spectrum in most of the other bins.
\begin{table}
\centering
\begin{tabular}[H]{l  c  c }
\hline
\hline\xrowht[()]{5pt}
Bias&$\chi_{S^3}^2$/d.o.f. \, (PTE)&$\chi^2_{T-RECS}$/d.o.f. \, (PTE)\cr
\hline
HB& 2.50 \, ($5.3\times 10^{-3}$)   &  4.88\, ($4.3\times 10^{-7}$) \cr 
THB&  1.85 \, ($4.7\times 10^{-2}$)    &  4.18 \, ($8.0\times 10^{-6}$) \cr 
PB&  1.59 \, ($1.0\times 10^{-1}$)   &  1.36 \, ($1.9\times 10^{-1}$)  \cr
TPB & 1.72 \, ($7.0\times 10^{-2}$) &  1.36 \, ($1.9\times 10^{-1}$) \cr 
 \hline
 \end{tabular}
\caption{Reduced $\chi^2$ values for the TGSS $\kappa g$ cross-spectrum and corresponding PTE values for all the bias models listed in column 1. Values in columns 2 and 3 refer to the T-RECS and $S^3$ $N(z)$ models, respectively. The number of degrees of freedom (d.o.f.) for this analysis is 10.
}
\label{tab:chi2_tgss}
\end{table}
Table \ref{tab:chi2_tgss} shows that these are in fact the models that provide the worst fit to the data. All the other models provide an acceptable fit to the data for $\ell>40$. The PB and TPB ones are those that perform better in both the T-RECS and $S^3$ cases. The $\chi^2$ analysis of the NVSS sample (results shown in Figure~\ref{fig:nvss_cross}) provides
similar results. All models but HB and THB with T-RECS number counts fail to match the amplitude of the cross-spectrum in the first $\ell$-bin. The analysis also confirms that the PB and TPB bias models perform better especially when coupled to the T-RECS $N(z)$ redshift distribution model. However, they generally provide worse PTE values than in the TGSS case, reflecting the smaller uncertainties of the measured NVSS $\kappa g$ cross-spectrum.

To summarize, the results of the cross-spectra analyses favor the PB and TPB bias models over the HB and THB ones, with a mild preference for the T-RECS $N(z)$ model over $S^3$ for the NVSS case. However, the PB and TPB models do not match the power amplitude on large angular scales. The significance of the mismatch is not high (about 2.4$\sigma$) but it is seen in both samples and, if confirmed, could indicate a genuine large scale excess power in the radio sources. 
\begin{table}
\centering
\begin{tabular}[H]{l  c  c }
\hline
\hline\xrowht[()]{5pt}
Bias&$\chi_{S^3}^2$/d.o.f. \,  (PTE)&$\chi^2_{T-RECS}$/d.o.f. \, (PTE)\cr
\hline
HB& 3.88 \, ($2.8\times 10^{-5}$)   &  17.55 \, ($0.00$) \cr 
THB&  3.32 \, ($2.6\times 10^{-4}$)    &  13.31 \, ($0.00$) \cr
PB&  2.46 \,  ($6.2\times 10^{-3}$)   &  2.08 \,  ($2.3\times 10^{-2}$)  \cr
TPB & 2.41 \, ($7.2\times 10^{-3}$) &  1.90 \, ($4.0\times 10^{-2}$) \cr 
 \hline
 \end{tabular}
\caption{Same scheme as for Table \ref{tab:chi2_tgss} but considering the NVSS $\kappa g$
cross-spectrum analysis.}
\label{tab:chi2_cross_nvss}
\end{table}

\subsection{Combined cross- and auto- angular power spectra results}
\label{sec:chi2_join}

The analysis of the $\kappa g$ cross-spectrum can only constrain the combination $b(z)\times N(z)$.
Since this degeneracy is potentially broken by combining the cross- and the auto-spectrum analysis, we perform a joint analysis using both. 
Considering that there is a significant evidence that the auto-spectrum of the TGSS catalog is affected by systematic errors, we restrict our joint analysis to the NVSS sample only. 

As explained in Section~\ref{sec:auto_spect}, we estimate the noise correction to the NVSS auto-spectrum, $N^{gg}$, from the data, since values from literature overestimate it. Also for the joint analysis, in the modelled auto-spectrum we include the noise term, $N^{gg}$, to account for both Poisson noise and spurious contribution from misidentified multiple sources. We treat $N^{gg}$ as a free parameter and estimate its value by minimizing the $\chi^2(N^{gg})$ function. The details of the procedure are described in Appendix \ref{sec:shot_noise_estimation} where we provide the best fit $N^{gg}$ values in Table~\ref{tab:shot_values} and also compare them with those obtained by using only the auto-spectrum.
We summarize the results of the $\chi^2$ joint analysis in Table~\ref{tab:chi2_joint_dl30}, where we list the minimum (with respect to $N^{gg}$) reduced $\chi^2$ values along with the corresponding PTE for all $N(z)$ and $b(z)$ models. Note that in this case the number of degrees of freedom is 13 (i.e. 14 elements of the data vector minus 1 fitted parameter).
\begin{table}[H]
\centering
\begin{tabular}[t]{l  c  c }
\hline
\hline\xrowht[()]{5pt}
Bias &$\chi_{S^3}^2$/d.o.f. \, (PTE)&$\chi_{T-RECS}^2$/d.o.f. \, (PTE)\cr
\hline
HB& 4.10 \, ($8.0\times10^{-7}$)   &  13.88   \, ($0.00$) \cr 
THB&  3.67  \, ($7.2\times10^{-6}$)    & 10.54   \,($0.00$) \cr 
PB&  2.34 \, ($4.1\times10^{-3}$)   &  1.93   \, ($2.2\times 10^{-2}$) \cr
TPB &  2.31   \, ($4.8\times10^{-3}$)&  1.79  \, ($3.9\times 10^{-2}$) \cr
 \hline
 \end{tabular}
\caption{
Same as Table \ref{tab:chi2_tgss} but for the combined NVSS $g g$ auto- and $\kappa g$ cross-spectra analysis. The number of degrees of freedom is 13.
}
\label{tab:chi2_joint_dl30}
\end{table}
Results of the joint analysis confirm those obtained using the cross-spectrum only. Bias models HB and THB are now even more strongly disfavored. The mild preference for using T-RECS in combination with PB and TPB with respect of using $S^3$ is also confirmed.



\section{Constraints on the effective bias of radio sources}\label{sec:bias_fit}

So far we tested the ability of existing $N(z)$ and $b(z)$ models to fit the observed auto $g g$ and cross $\kappa g$ spectra, with no attempt to set constraints on either models. 
In this Section, we change strategy and we use the spectra to constrain the effective bias of the radio sources and its evolution, $b(z)$, having assumed a model for their redshift distribution $N(z)$.
We do not adopt the alternative strategy of assuming $b(z)$ while leaving $N(z)$ free to vary since the observational constraints of the redshift distribution of the radio galaxies are tighter than those for their bias.
In any case, to account for the theoretical uncertainties in the $N(z)$ model we consider both T-RECS and $S^3$ and propagate this difference to the $b(z)$ uncertainty.

We consider two bias models, dubbed \textit{model 1} and \textit{model 2}, that have been introduced in Section~\ref{sec:theory} and depend on one free parameter only, $b_g$. 
Their simplicity is motivated by the need of minimizing the free parameters of the model since the number of data points in the analysis is limited (10 and 14, in the cross- and joint-spectra analyses, respectively). 
To estimate $b_g$, we search for the minimum value, $\chi^2_{min}$, of the $\chi^2(b_g)$ function evaluated in 500 equally spaced points in the interval $[b_{min}, b_{max}]$. 
The angular spectrum model is evaluated in each of the 500 points whereas the covariance matrix is set equal to that of the "fiducial" model which is provided by the best fit found in Section~\ref{sec:modelvsdata}. Therefore, we adopt the one based on the TPB bias model and on either the $S^3$ or the T-RECS $N(z)$, depending on which one is used for the angular spectrum model. Following the Gaussian hypothesis, we estimate the 1$\sigma$ uncertainty of the $b_g$ parameter by setting $\Delta \chi^2 = \chi^2(b_g)- \chi^2_{min}= 1$.

Like in the previous Section, we first perform the $\chi^2$ minimization using the cross-spectrum information only. In this case we consider both the TGSS and the NVSS sample.
Then we repeat the procedure using both the cross- and the auto-spectra. In this second case we restrict the analysis to the NVSS case only.

\begin{figure}[!t]
    \centering
    \includegraphics[width = \hsize]{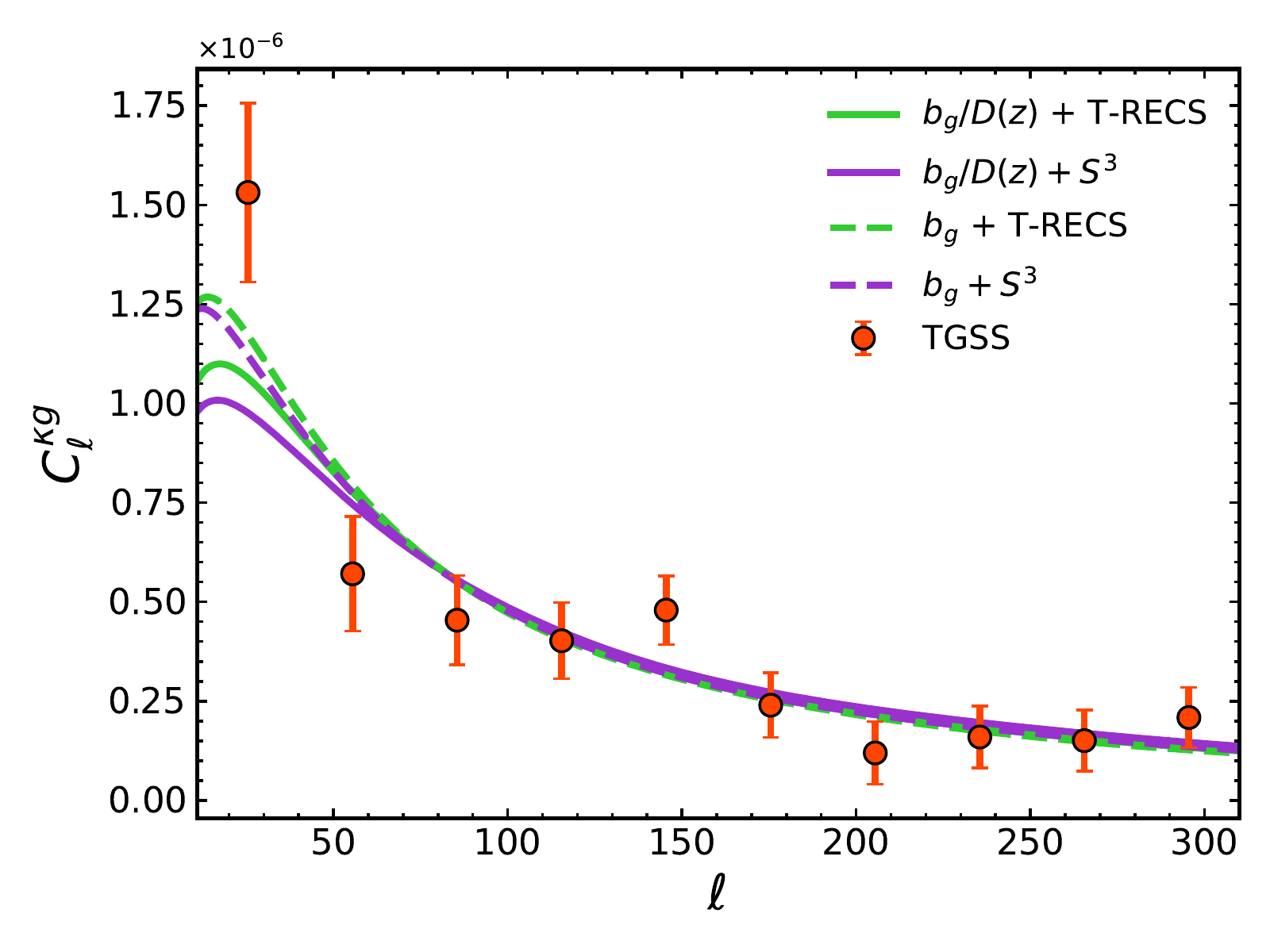}
    \caption{Best fit $\kappa g$ power spectra models for the TGSS sample (solid and dashed lines) vs. measured (red dots) and their 1$\sigma$ Gaussian error bars. 
    Different colors and line styles are used for the different models. Light green: $N(z)$ from T-RECS. Purple: $N(z)$ from $S^3$. Continuous: \textit{model 1}, linearly evolving bias. Dashed: \textit{model 2}, constant bias.
    The best fitting $b_g$ values are listed in Table~\ref{tab:tgss_kappa_bias_bf}.}
    \label{fig:tgss_kappa_bias_bf}
\end{figure}

\begin{figure}
    \centering
    \includegraphics[width=\hsize]{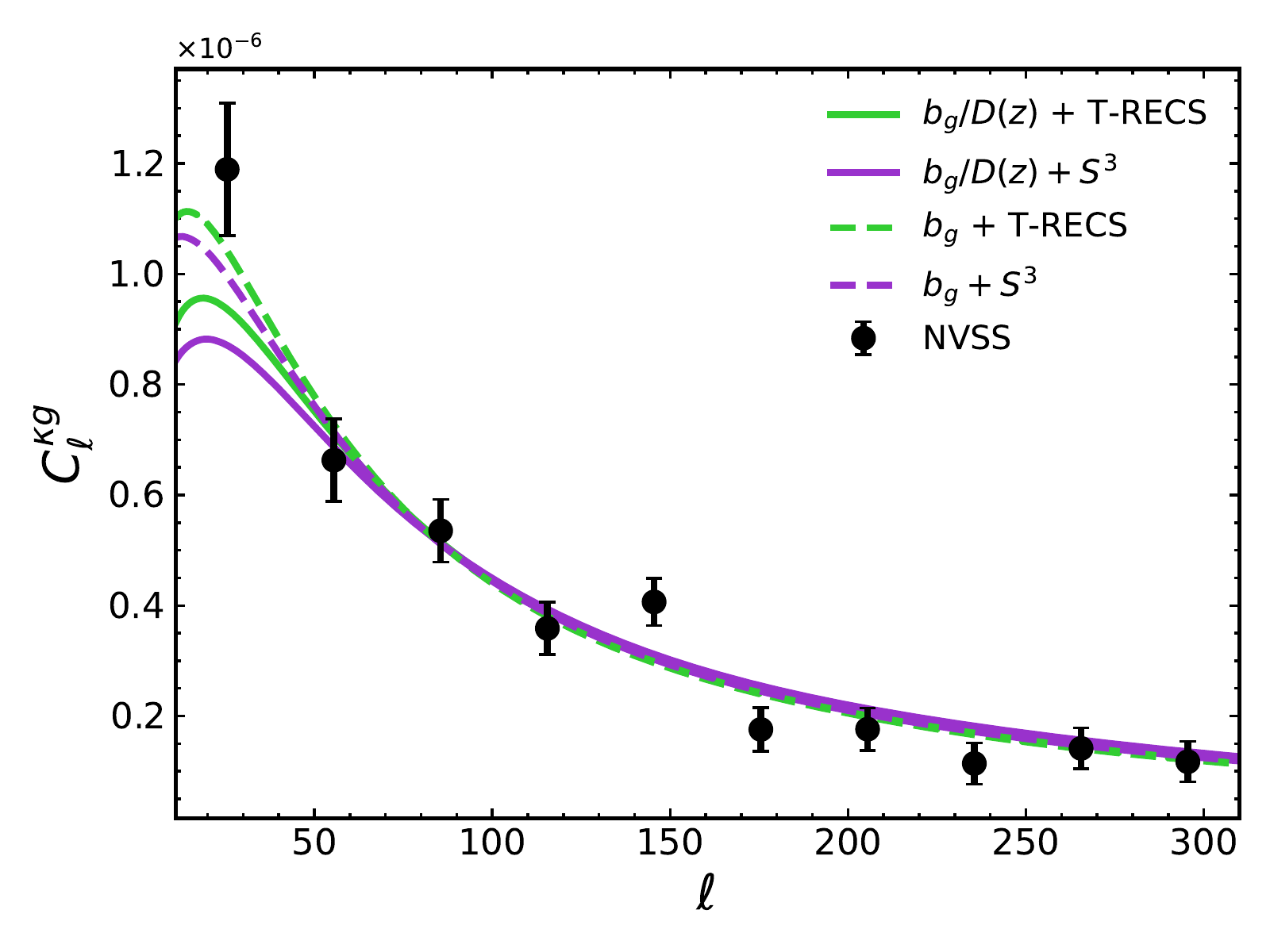}
    \caption{Same as Figure~\ref{fig:tgss_kappa_bias_bf} but considering the NVSS sample. Data are now shown with black diamonds along with their $1\sigma$ errorbars. Colors and line styles of the model cross-spectra are also the same. Values of the best fit $b_g$ are summarized for the different models in Table~\ref{tab:nvss_kappa_bias_bf}.}
    \label{fig:nvss_kappa_bias_bf}
\end{figure}

For the TGSS catalog, results are shown in Figure~\ref{fig:tgss_kappa_bias_bf} while the best fitting bias parameters are listed in Table~\ref{tab:tgss_kappa_bias_bf}.
For NVSS results are instead summarized in Figure~\ref{fig:nvss_kappa_bias_bf} and Table~\ref{tab:nvss_kappa_bias_bf}.

\begin{table}[H]
\centering
\begin{tabular}[t]{l  c  c  }
\hline
\hline\xrowht[()]{5pt}
N(z) & $b_g \pm 1 \sigma$ & $\chi^2$/d.o.f. \, (PTE)\cr
\hline
T-RECS& $ 1.72/D(z) \pm 0.17$&  1.52  \, ($1.33\times 10^{-1}$) \cr 
T-RECS &  $2.59 \pm 0.21$& 1.36  \, ($2.00\times 10 ^{-1}$) \cr 
$S^3$&  $1.87/D(z) \pm 0.18 $&  1.67  \, ($9.01\times 10^{-2}$) \cr 
$S^3$& $2.99\pm 0.24$&  1.39  \, ($1.86\times10^{-1}$) \cr
 \hline
 \end{tabular}
\caption{Best fit $b_g$ values with 1$\sigma$ Gaussian errors (column 2) for the various $N(z) $ models (column 1) used in the TGSS $\kappa g$ cross-spectrum analysis. The reduced $\chi^2_{min}$ values (estimated for 9 degrees of freedom)
and the corresponding PTE are given in column 3.
}
\label{tab:tgss_kappa_bias_bf}
\end{table}

\begin{table}[H]
\centering
\begin{tabular}[t]{l  c  c  c }
\hline
\hline\xrowht[()]{5pt}
N(z) & $b_g \pm 1 \sigma$ & $\chi^2$/d.o.f. \, (PTE)\cr
\hline
T-RECS& $1.64/D(z)\pm 0.09$ & 2.07  \, ($2.85\times 10^{-2}$) \cr
T-RECS &  $2.53 \pm 0.12 $ & 1.59  \, ($9.93\times 10 ^{-2}$) \cr 
$S^3$&  $1.73/D(z) \pm 0.10$ &  2.48  \, ($7.92\times 10^{-3}$) \cr
$S^3$& $2.79 \pm 0.12$ &  1.71  \, ($6.98\times10^{-2}$) \cr 
 \hline
 \end{tabular}
\caption{Same as Table~\ref{tab:tgss_kappa_bias_bf} but for the NVSS catalog.}
\label{tab:nvss_kappa_bias_bf}
\end{table}

The $b_g$ values of TGSS  are consistent, though somewhat higher than those found for the NVSS sample. This is expected since TGSS objects are typically brighter, and therefore more biased, than the NVSS ones. The fact that the PTE values for TGSS are smaller than for NVSS 
simply reflects the fact that errors in the measured $\kappa g$
TGSS spectrum are larger than in the NVSS one because of the smaller number of sources.

Leaving the bias model free to vary generally improves the quality of the fit and significantly reduces the mismatch between models and data in the first multipole bin. Moreover, the constant \textit{model 2} bias fits the data better than the evolving \textit{model 1}, irrespective of the $N(z)$ model adopted. In fact, for any given bias model, 
the two best fit $b_g$ values obtained using either T-RECS or $S^3$ are consistent with each other, indicating that these results are rather insensitive to the $N(z)$ model uncertainty. On the contrary, the $b_g$ value is sensitive to the bias model adopted, being systematically larger for \textit{model 1} than for \textit{model 2}.

\begin{figure*}
\begin{multicols}{2}
    \includegraphics[width=\linewidth]{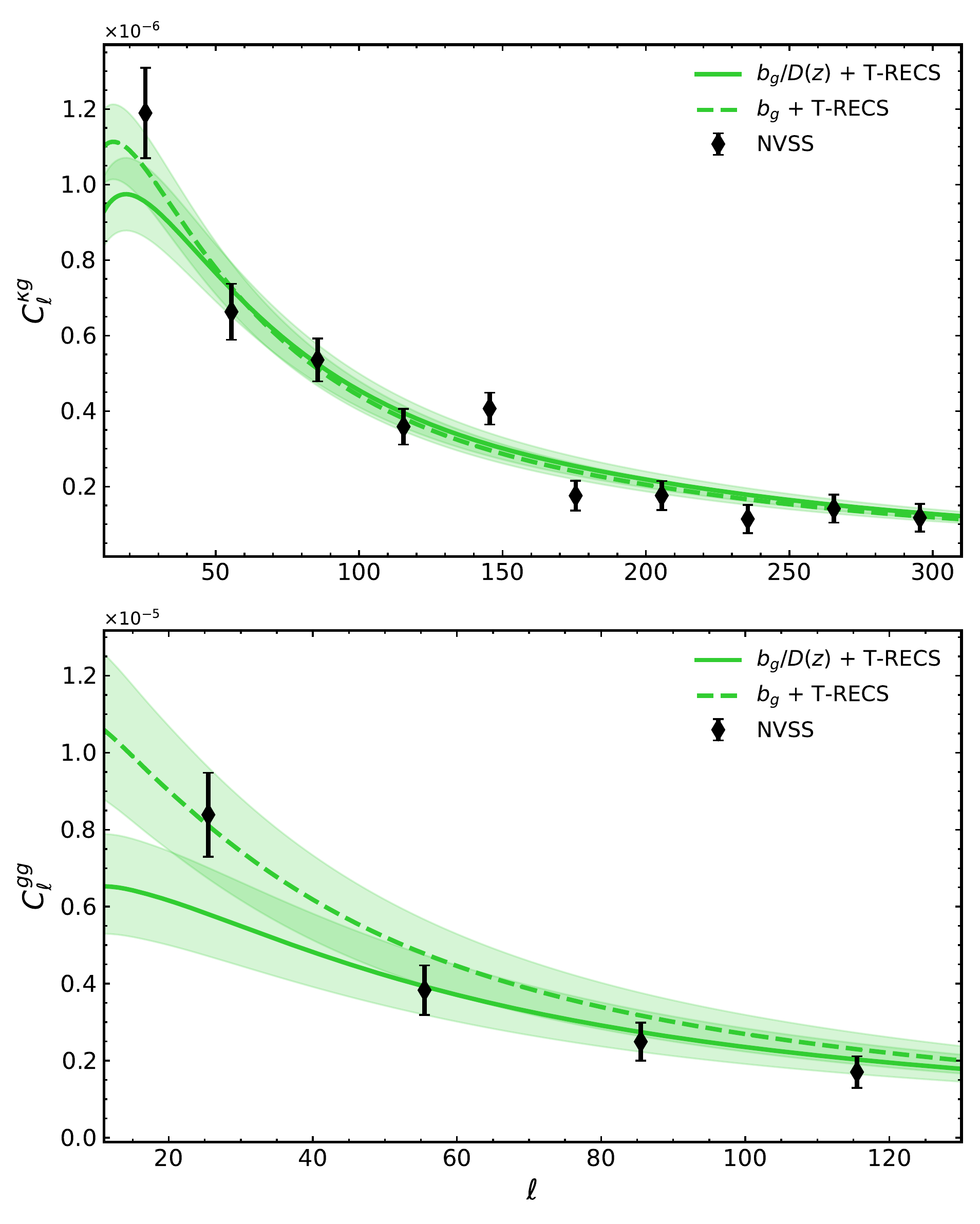}\par
    \includegraphics[width=\linewidth]{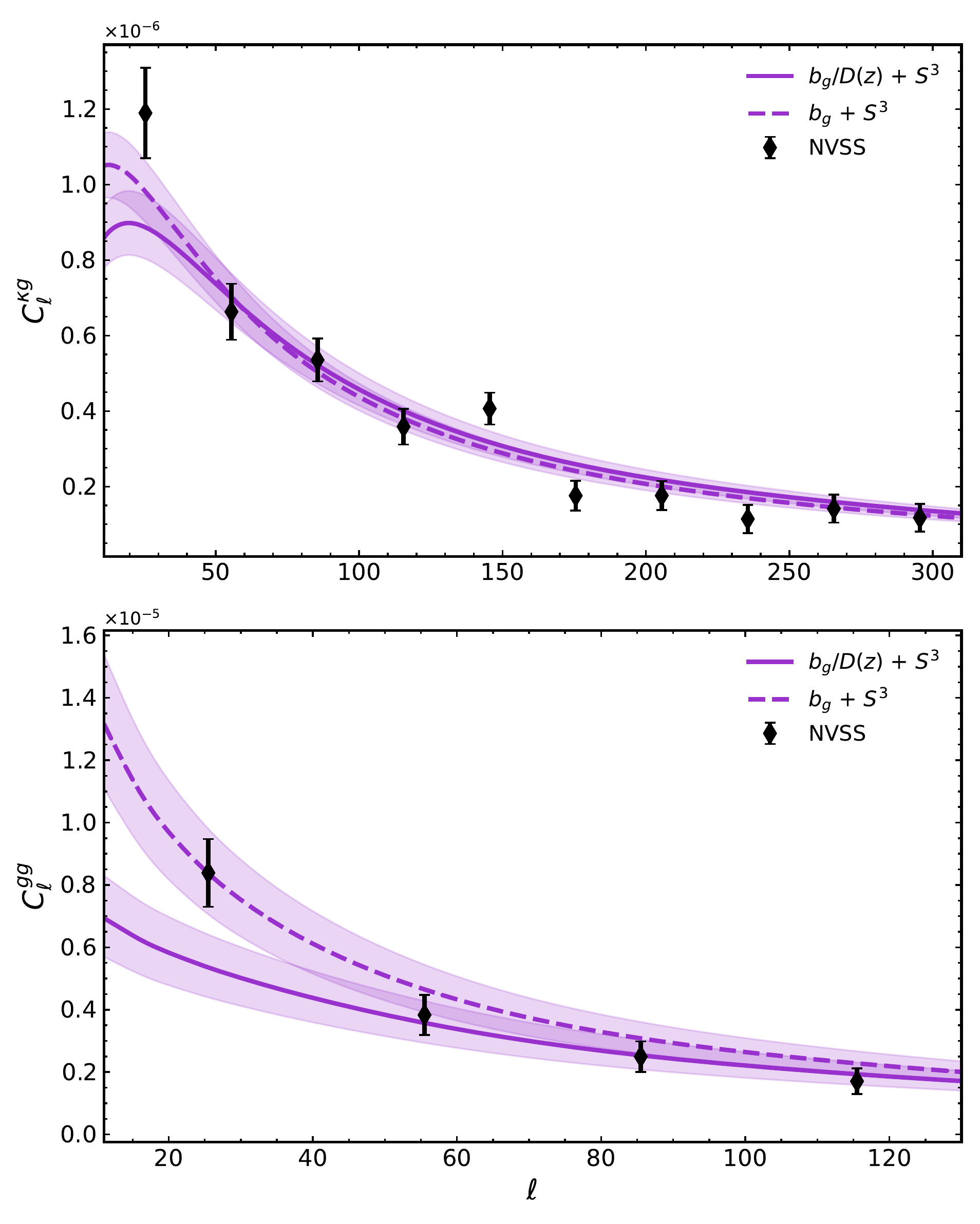}\par
\end{multicols}
    \caption{Model vs. measured NVSS $\kappa g$ (top panels) and $gg$ (bottom panels)
    angular power spectra. The measured cross-spectra (top panels) are the same as in the previous plots. Errorbars represent 1$\sigma$ Gaussian uncertainties.
    The curves show model predictions. They are surrounded by shaded areas that represent the 2$\sigma$ uncertainty interval of the best fitting $b_g$ values.
    The green color in the left panels indicates models that adopt the T-RECS $N(z)$ prescription.
    The purple color flags the adoption of the $S^3$ model in the right panels.
    Continuous and dashed lines are used for bias \textit{model 1} and \textit{model 2}, respectively. 
    All best fitting $b_g$ values are listed in Table~\ref{tab:nvss_kappa_bias_bf_joint}.}
    \label{fig:nvss_kappa_joint_fit_2sigma}
\end{figure*}

For the joint analysis, we also consider the four additional data points of the NVSS $gg$ auto-spectrum. One important difference with respect to the cross-spectrum analysis only is the fact that we now minimize the $\chi^2$ function with respect to two free parameters: the effective bias $b_g$ and the auto-spectrum noise term $N^{gg}$. We therefore search for the minimum of the $\chi^2_{2D}(b_g,N^{gg})$ function. 
To find the 1$\sigma$ confidence level of each parameter, we firstly estimate the 2D joint probability function, $P_{2D} = e^{-\chi^2_{2D}/2}$. Then we marginalize over one parameter to obtain the 1D probability function of the other, $P_{1D}$, and consider its $16th$ to $84th$ percentile interval.
\begin{figure}
    \centering
    \includegraphics[width = \hsize]{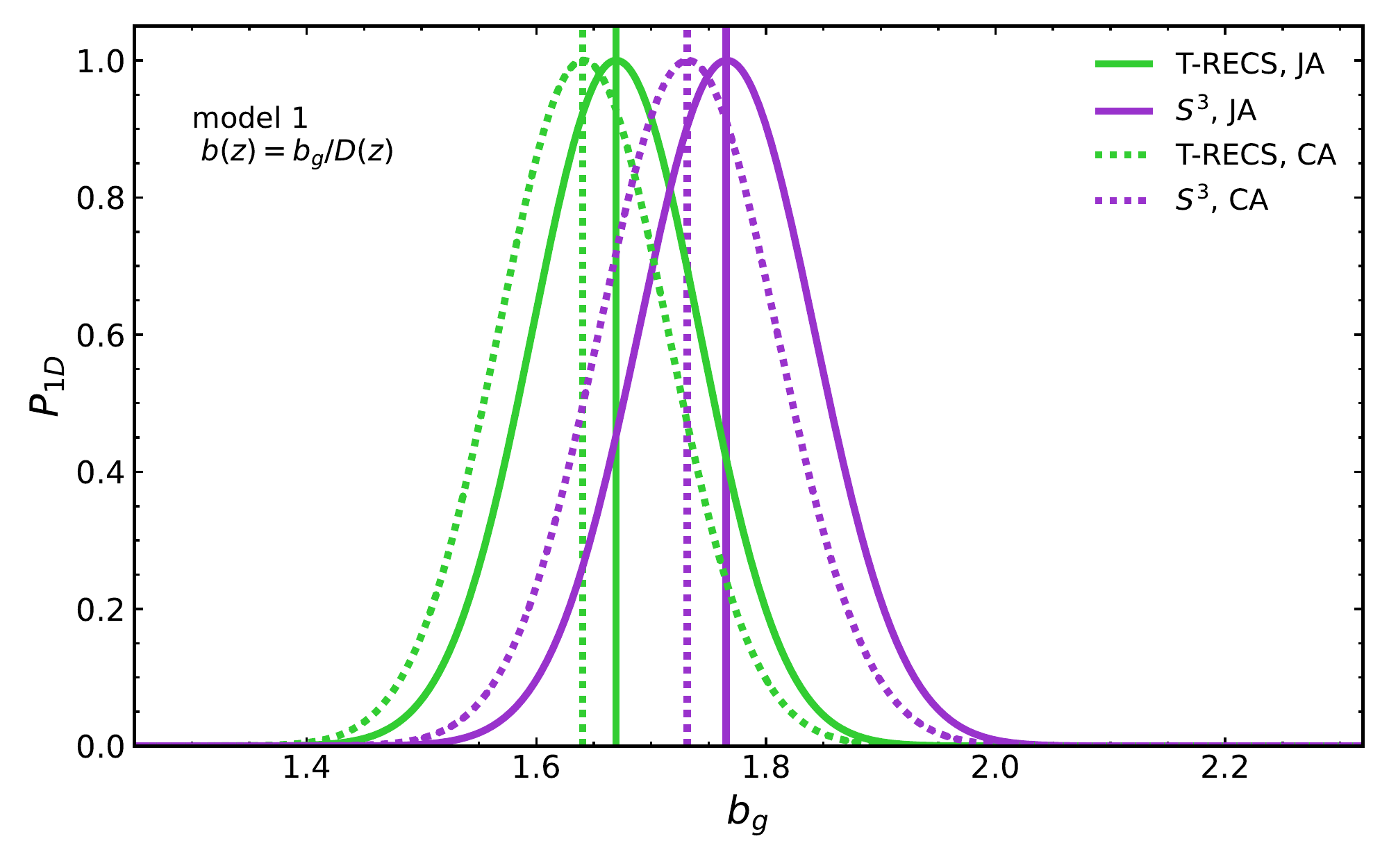}\par
    \includegraphics[width = \hsize]{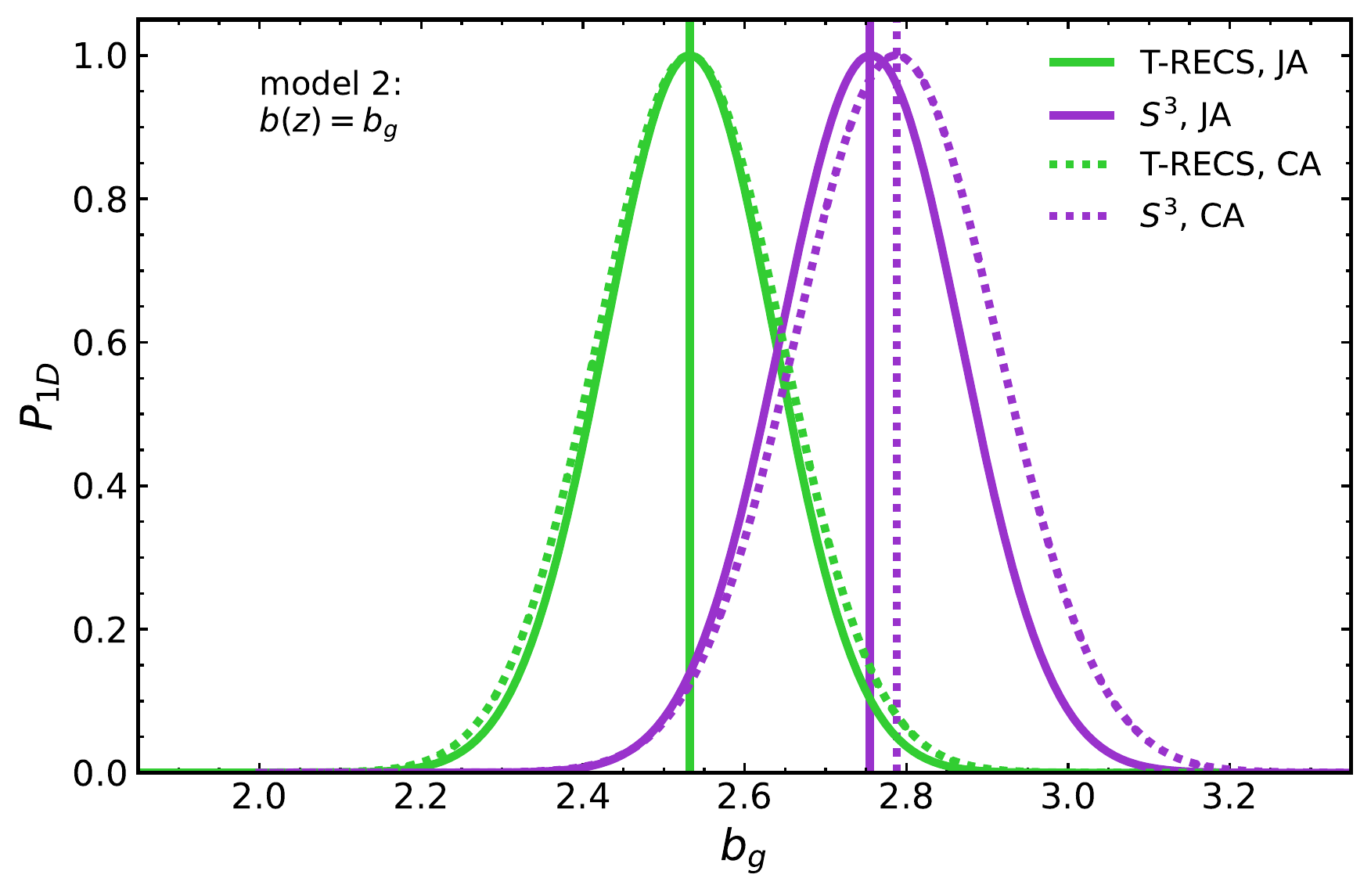}
    \caption{The continuous curves show the marginalized 1D probability distributions, $P_{1D}$, of $b_g$ for \textit{model 1} (constant bias) and \textit{model 2} (linearly growing bias), obtained from the joint NVSS $gg$ and $\kappa g$ power spectrum analysis. All curves are normalized so that $P_{1D}=1$ at the maximum. The different colors indicate the different $N(z)$ models and match those used in Figure~\ref{fig:nvss_kappa_bias_bf}.
    Dotted curves are drawn for comparison and show the 1D probability distribution from the NVSS cross-spectrum only analysis. They are labelled CA to further distinguish them from those obtained from the joint analysis, labelled JA.}
    \label{fig:nvss_joitn_bf_trecs_bconst_bDz}
\end{figure}
The 1D probability distributions for $b_g$ obtained for the joint analysis (continuous curves in Figure~\ref{fig:nvss_joitn_bf_trecs_bconst_bDz}) are very similar to those obtained from the NVSS cross-spectrum only analysis (dotted curves).
The best fitting $b_g$ values found in correspondence of the maxima are indicated by a vertical line and listed in Table~\ref{tab:nvss_kappa_bias_bf_joint} along with their 1$\sigma$ errors. In the Table, we also show the value of the noise term $N^{gg}$ with its uncertainty and the minimum value of the reduced $\chi^2_{2D}$ along with its PTE. Note that the estimates of $N^{gg}$ are in agreement with those computed in Appendix~\ref{sec:shot_noise_estimation} for the bias models taken from the literature.
The results of the joint analysis agree with those obtained from the cross-spectrum only and confirm their weak sensitivity to the choice of the $N(z)$ model. However, unlike in the joint analysis presented in Section~\ref{sec:chi2_join}, the addition of the auto-spectrum does change the results since the preference for bias \textit{model 2} over \textit{model 1} which was rather weak in the $\kappa g$ only analysis, it is now significantly stronger. The reason for this is clearly seen in  Figure~\ref{fig:nvss_kappa_joint_fit_2sigma} in which we compare both the NVSS $\kappa g$ (top panels) and $ g g$ (bottom panels) measured spectra (black diamonds) with the different best fitting models (continuous solid and dashed curves). 
Shaded areas represent the 2$\sigma$ uncertainty interval of the estimated $b_g$ parameter. 
The two model cross-spectra, whose amplitude scales as $b_g$, provide similar predictions. This is not the case for the auto-spectra for which the two models, whose amplitudes scale as $b_g^2$, significantly depart from each other at small $\ell$ values.
 The preference of bias \textit{model 2} over bias \textit{model 1} largely stems from this different behavior.
 The quality of the fit has also improved, especially for the bias \textit{model 2} case, with respect to the $\kappa g$ only analysis as indicated by the larger PTE values.

\begin{table*}
\centering
\begin{tabular}[t]{l  c  c  c }
\hline
\hline\xrowht[()]{5pt}
N(z) & $b_g \pm 1 \sigma$ & $N^{gg} \pm 1\sigma$ & $\chi^2$/d.o.f. \, (PTE)\cr
\hline
T-RECS & 1.67/D(z) $\pm \, 0.09$ & (1.88$\, \pm \,0.03)\times10^{-5}$& 1.87 \, ($3.25\times 10^{-2}$) \cr 
T-RECS &  2.53 $\pm \, 0.11$ & (1.82$\, \pm \, 0.04)\times 10 ^{-5} $& 1.24 \, ($2.51\times 10 ^{-1}$) \cr 
$S^3$ &  1.76/D(z) $\pm \, 0.10$ & (1.90$\, \pm \, 0.03)\times10^{-5}$& 2.20 \, ($9.30\times 10^{-3}$) \cr 
$S^3$ & 2.75 $\pm \, 0.11$ & (1.82$\, \pm \, 0.04) \times 10 ^{-5}$& 1.36 \, ($1.79\times 10^{-1}$) \cr 
 \hline
 \end{tabular}
\caption{Best-fit values for the NVSS bias parameter $b_g$ and for the auto-spectrum noise term $N^{gg}$ obtained from the joint $\kappa g$ and $ gg$ angular spectra analysis.
Both parameters are listed with their 1$\sigma$ Gaussian uncertainties. The reduced $\chi^2$ value (12 degrees of freedom) at its minimum and the corresponding PTE are also shown in the last column.
}
\label{tab:nvss_kappa_bias_bf_joint}
\end{table*}

\section{Robustness tests}
\label{sec:robustness_tests}
In this Section we present a number of tests designed to assess the robustness of the results of the cross-spectrum analyses presented so far.
We focus on the cross-spectra results only for two reasons. First, constraints on the $N(z)$ and bias models largely come from this statistics alone. Secondly, similar robustness tests have been performed, successfully, on the TGSS and NVSS auto-spectra by \cite{dolfi_2019}.
These tests are designed to assess the impact of possible observational systematic effects on either the radio catalogs or the convergence maps, or both.

\subsection{Robustness to residual astrophysical foreground contamination}
\label{sec:galcuts}

To minimize the impact of foreground Galactic emission and that of confusion associated to high stellar density,
in Section~\ref{sec:data} we masked out the sky area that is close to the Galactic plane, both in the CMB lensing convergence map and in the TGSS and NVSS source catalogs.

\begin{figure}
    \centering
    \includegraphics[width=\hsize]{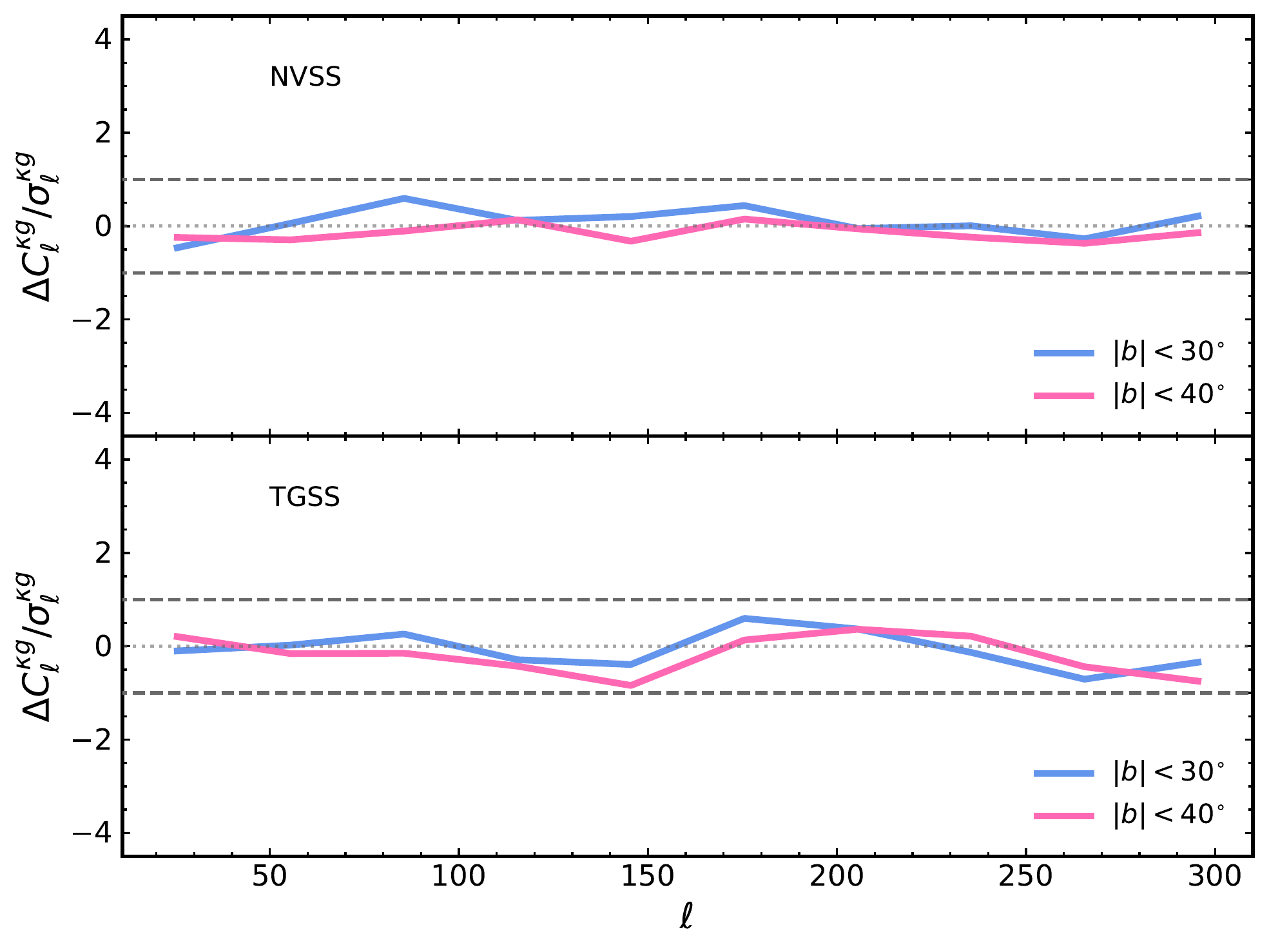}
    \caption{Residuals $\Delta C_{\ell}^{\kappa g}$ between cross-spectra estimated with the baseline mask and those computed with the more conservative masks featuring Galactic latitude cuts $|b|<30^{\circ}$ (blue curve) and  $|b|<40^{\circ}$ (magenta). Residuals on the y-axis are in units of the Gaussian random errors, whose 1-$\sigma$ level is plotted for reference as a horizontal dashed line. Top and bottom panels show the results for the NVSS and TGSS cases.}
 \label{fig:nvss_kappag_bcuts}
\end{figure}

To assess the robustness of our analysis to the choice of the mask, we repeated the cross-correlation analysis using two more aggressive Galactic cuts to exclude regions with $|b|< 30^{ \circ}$ and $|b|< 40^{\circ}$ in all maps. As a result, the sky fraction considered in the analysis decreases to $f^{\kappa g}_{\rm sky,30^{\circ}} = 0.40$ $(0.41)$ and $f^{\kappa g}_{\rm sky,40^{\circ}} = 0.29$ $(0.30)$ for the cross-correlation with NVSS (TGSS). 
We compare the results with those obtained with the baseline sky masks of Section~\ref{sec:data}. In Figure~\ref{fig:nvss_kappag_bcuts} we show the difference between the new and the baseline cross-spectra, $\Delta C_{\ell}^{\kappa g}$, in units of the Gaussian error $\sigma_{\ell}^{\kappa g}$ from Equation~\ref{eq:diag_cov} estimated
considering the TPB bias model, the T-RECS model for $N(z)$ and using the most aggressive mask. The upper and bottom panels show the results for the NVSS and TGSS cross-spectra, respectively. Horizontal dashed  lines are drawn at the 1$\sigma$ level value for reference. For both mask choices and for both radio source catalogs, the residuals show no significant trends with $\ell$ or $f_{\rm sky}$ and their amplitude oscillates within the dashed lines. We conclude that the results of our analysis, 
obtained with the baseline mask, are robust to possible systematic effects associated to Galactic foreground contamination.

\begin{figure}
    \centering
    \includegraphics[width=\hsize]{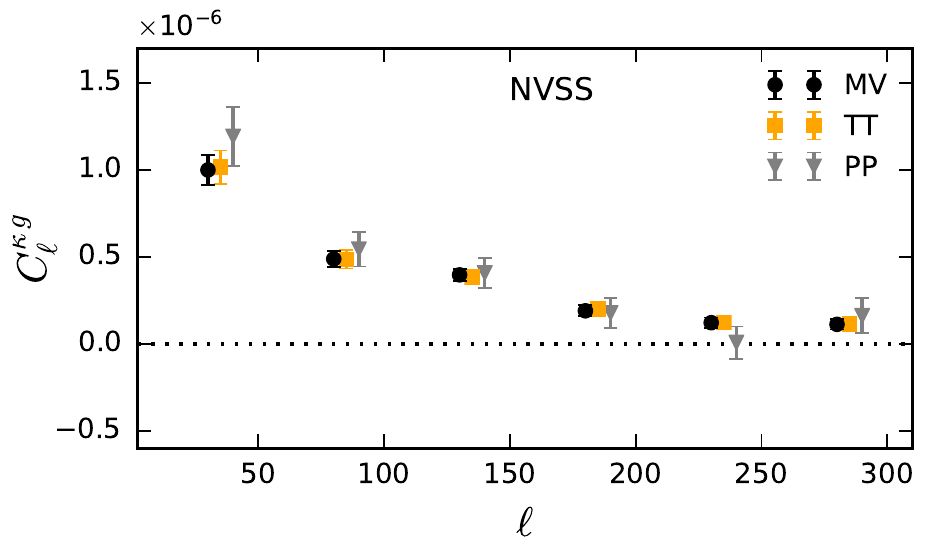}
    \includegraphics[width=\hsize]{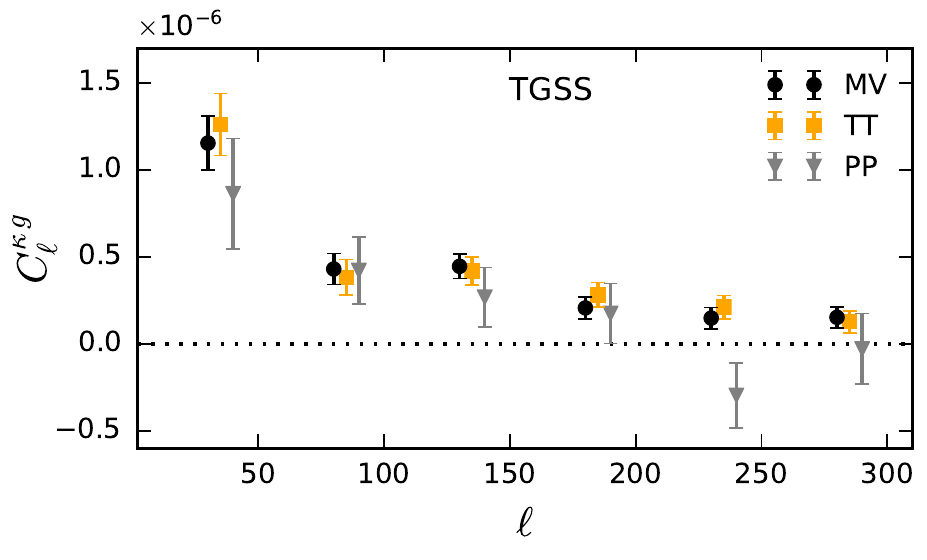}
    \caption{Cross-angular power spectra for NVSS (top) and TGSS (bottom) with different Planck CMB lensing reconstructions obtained with temperature only data (TT), polarization only data (PP) and their minimum variance combination (MV). To estimate the spectra we use the baseline mask and a multipole bin size $\Delta_{\ell} = 50$.}
    \label{fig:lensing_TTPPMV}
\end{figure}

Besides Galactic foregrounds, another possible source of contamination is represented by extragalactic point sources. Extragalactic objects are expected to trace the same LSS as the radio sources. Moreover, since they follow a highly non-Gaussian distribution, they could bias the CMB lensing maps since the lensing reconstruction relies on the non-Gaussian nature of the lensed CMB. In particular, contamination from extragalactic point sources can even correlate with the radio sources distribution we are investigating.
As a result, they can potentially bias the cross-correlation spectrum. To assess their impact on our cross-correlation analysis, we consider a different lensing reconstruction, that obtained from polarization CMB data. 
At present, the extragalactic contamination in polarization is not robustly quantified, but it is expected to be less of a problem because of the comparative lower fraction of polarized sources \citep{CMBPOL}. While the Planck lensing is mostly dominated by the CMB temperature information, an independent lensing map has been generated and only accounts for CMB polarization data \citep{planck_lens}. In Figure \ref{fig:lensing_TTPPMV}, we compare the angular power spectra estimated by cross-correlating NVSS and TGSS sources with Planck lensing reconstructions obtained from CMB temperature only data (TT), polarization only data (PP) and their minimum variance combination (MV), the latter being the baseline lensing map we used in this work. The polarization reconstruction is significantly noisier, an aspect which motivated us to 
use wider multipole bins $\Delta_{\ell} = 50$ for this comparison, instead of the baseline one $\Delta_{\ell} = 30$.
Figure \ref{fig:lensing_TTPPMV} shows that different lensing reconstructions provide compatible results. Hence, we conclude that systematic errors induced by extragalactic sources contamination are well within the expected statistical uncertainties of our analysis.

\subsection{Robustness to radio flux cut}
\label{sec:fluxcuts}

As we discussed in Section \ref{sec:intro}, wide radio surveys are prone to large scale variations in the flux calibration. This effect can modulate the completeness of the sample and generate spurious signals in the angular spectra at low multipoles.
The amplitude of the effect is expected to be larger at fainter fluxes, where the completeness of the catalog drops. To assess the potential impact of this effect we follow \cite{dolfi_2019} and select sub-samples of NVSS and TGSS radio sources using different (lower) flux cuts $ S_{min}$ that we cross-correlate with the CMB convergence maps.
 
\begin{figure}
    \centering
    \includegraphics[width=\hsize]{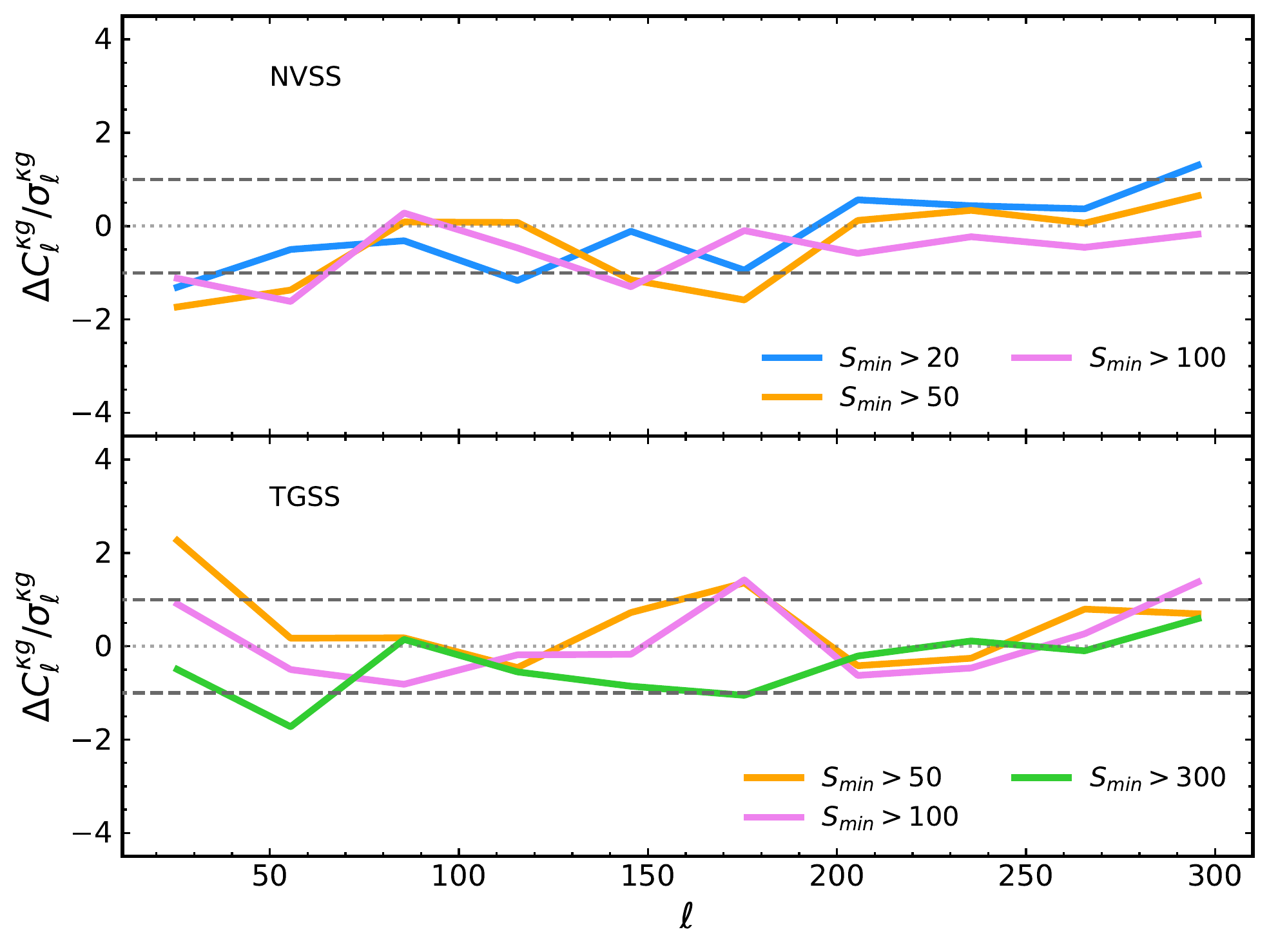}
    \caption{Residuals $\Delta C_{\ell}^{\kappa g}$ in units of 1$\sigma$ Gaussian errors obtained using different NVSS (top) and TGSS (bottom) sub-catalogs selected at different flux cuts
    $ S_{min}$. Different colors are used for the different choices of $ S_{min}$ indicated in the plots. The dashed horizontal lines bracket the 1$\sigma$ Gaussian uncertainty region. Residuals are estimated with respect to the baseline $C_{\ell}^{\kappa g}$ as described in the text.
}
    \label{fig:tgss_nvss_kappag_fcuts}
\end{figure}

For the NVSS case we gradually increase the flux cut from the baseline value  $S_{min}= 10 \ \SI{}{\mJy}$ up to $20, \, 50\; \rm{and}\; 100 \ \SI{}{\mJy}$.
The results are shown in the top panel of Figure~\ref{fig:tgss_nvss_kappag_fcuts},  in which we plot the difference of the $\kappa g$ cross-correlation spectra with respect to the baseline case, in units of the 1$\sigma$ Gaussian error. The latter is the same as the baseline Gaussian error in which, however, the Poisson noise term is the one of the sub-catalog selected at the new $S_{min}$ value.
The normalized residuals are confined within the 1$\sigma$ uncertainty strip, which indicates that the results of the NVSS cross-correlation analysis are robust. Since a similar robustness was found for the NVSS auto-correlation analysis \citep{dolfi_2019}, we conclude that also the joint analysis is robust to cutting the sample at $S_{min} \geq 10 \ \SI{}{\mJy}$.

We repeat the same analysis for the TGSS case and find similar results.
Here we consider cuts at $ S_{min}= 50$ and $100 \ \SI{}{\mJy}$ that are less aggressive than the baseline case ($200 \ \SI{}{\mJy}$) as well as a more conservative one for which $ S_{min}= 300 \ \SI{}{\mJy}$.
As shown in Figure~\ref{fig:tgss_nvss_kappag_fcuts}, residuals are generally below the 1$\sigma$ Gaussian error, which indicates that the TGSS cross-correlation analysis is robust to the choice of $ S_{min}$.

Finally, we also checked that results are insensitive to the choice of the upper flux cut, $S_{max}$, as in the case of the auto-correlation analysis \citep{dolfi_2019}. We consider two different values, $S_{max} = 3000, 5000 \ \SI{}{\mJy}$. The residuals, computed with respect to the baseline value $S_{max}= 1000\ \SI{}{\mJy}$, show no significant trend with the value of multipoles or flux cuts and we conclude that the cross-correlation is robust also to the changing of $S_{max}$.

\subsection{Robustness to lensing magnification bias modelling}\label{sec:magbias}

\begin{figure}
    \centering
    \includegraphics[width=\hsize]{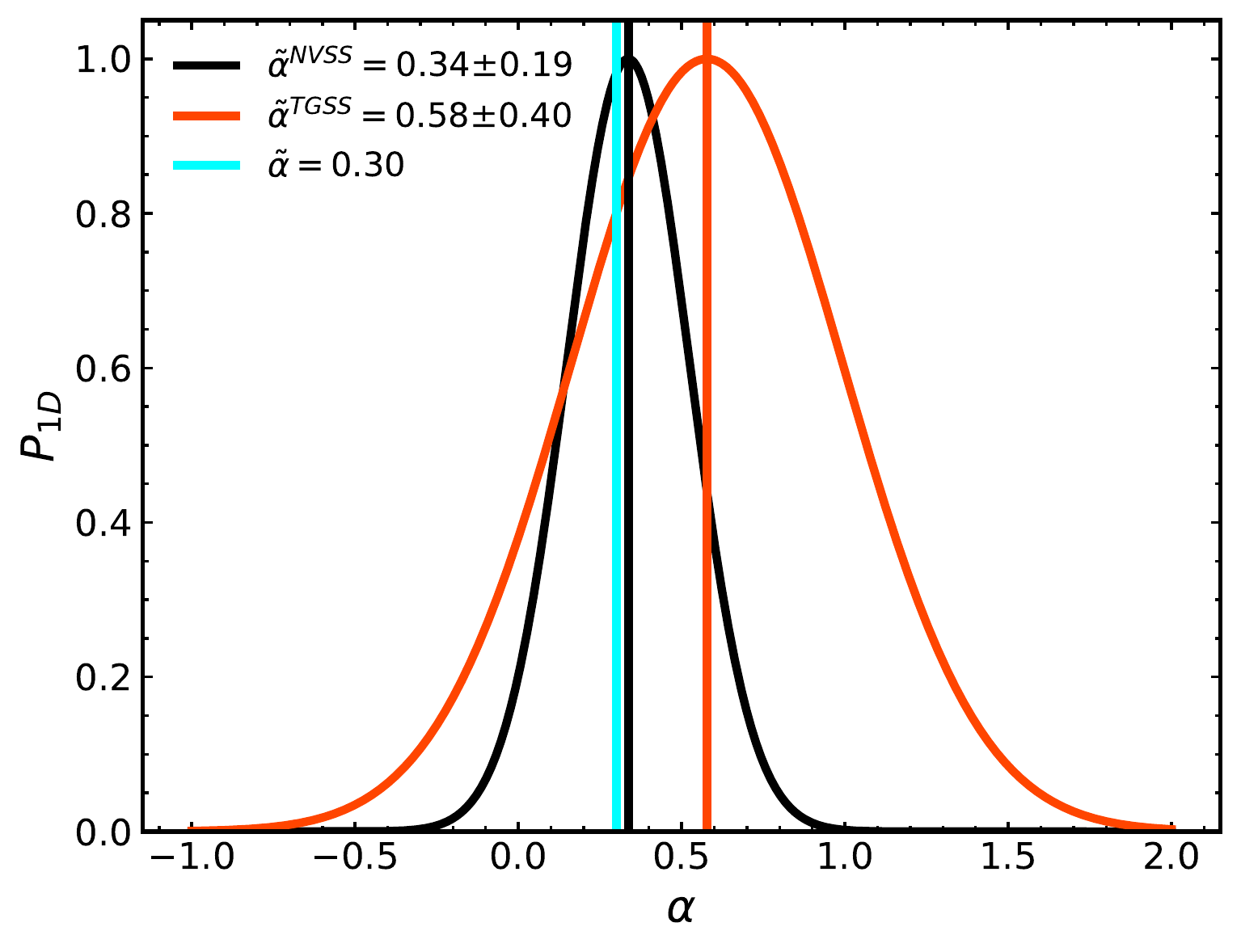}
    \caption{Probability distribution functions for the magnification bias parameter $\tilde{\alpha}$ obtained from the NVSS (black curve) and TGSS (red curve) $\kappa g$ cross-spectra analysis. The probabilities are normalized with respect to the maximum. Positions of the maxima are flagged by vertical lines. The cyan line indicates the baseline $\tilde{\alpha} = 0.30$ value used throughout this work.}
    \label{fig:prob_1d_alpha}
\end{figure}

In the angular spectra models introduced in Section~\ref{sec:theory}, we account also for the lensing magnification bias, which has an impact on large angular scales. The effect is fully quantified by a single parameter $\tilde{\alpha}$, which represents the effective slope of the faint end luminosity function of the radio sources. It is defined as the weighted mean of the individual luminosity function slopes of the various objects' types (Equation~\ref{eq:magnification_bias}). 
So far, we use $\tilde{\alpha} = 0.30$ i.e. we set this value equal to that of \cite{dolfi_2019} originally evaluated for TGSS and assuming a $S^3$ $N(z)$ model.
However, the value of the $\tilde{\alpha}$ depends on the type of objects included in the catalog and on their individual redshift distributions.
To investigate the sensitivity of our results to the choice of $\tilde{\alpha}$, we repeat the cross-correlation analysis for both the TGSS and NVSS cases using a T-RECS $N(z)$ model and assuming a TPB bias while the $\tilde{\alpha}$ parameter is free to vary. Then, we search for the minimum of the $\chi^2(\tilde{\alpha})$ function evaluated in 600 equally spaced points in the range $\tilde{\alpha} = [-1.0,2.0]$. 
The corresponding 1D probability distribution functions for $\tilde{\alpha}$ 
are shown in Figure~\ref{fig:prob_1d_alpha} and compared with the reference value $\tilde{\alpha} = 0.30$ (cyan vertical line).
The two probability functions peak at two different $\tilde{\alpha}$ values, 
confirming that $\tilde{\alpha}$ does depend on the catalog used. Moreover, the difference between the TGSS best fit value $\tilde{\alpha} = 0.58$ and that of \citet{dolfi_2019} quantifies the sensitivity to the $N(z)$ model.
That said, the differences between the reference and the best fit $\tilde{\alpha}$ values is within the variance of the two distributions, both close to Gaussian. We therefore conclude that our analysis is robust also to the choice of $\tilde{\alpha} = 0.30$.

\section{Discussion and conclusions}
\label{sec:conclusion}

In this work we investigated the cross-correlation between wide surveys of extragalactic radio sources, namely TGSS and NVSS, and CMB lensing. Both are tracers of the underlying mass distribution and potentially affected by different and supposedly uncorrelated observational systematic uncertainties. Their cross-correlation analysis should then be insensitive to those systematic errors that might instead affect the auto-correlation measurements. 
Moreover, a joint analysis that combines auto- and cross-correlation statistics is able to break, at least in part, the degeneracy between the bias and the redshift distribution of the tracers, i.e. two quantities that are weakly constrained in the case of wide radio surveys.

The main results of our analysis are the following:
\begin{itemize}
\item We confirm an excess clustering power of the TGSS sample over the one from NVSS at large angular scales. This excess has been detected with high statistical significance in previous analyses in the multipole range $\ell \leq 40$ and regarded as a spurious feature that should be attributed to some observational or instrumental effect \citep{Tiwari_2019}. 
When the same flux cuts, sky areas and $\ell-$ binning are adopted, we reproduce the results of \cite{dolfi_2019} which constitutes a consistency test for our analysis pipeline.

\item The cross-correlation spectrum of the CMB lensing convergence - NVSS is in good agreement with the CMB lensing convergence - TGSS one in the range $\ell=[11, 310]$. 
The choice of excluding multipoles $\ell<11$ is conservative and motivated by the known dependence of the flux sensitivity on the Galactic latitude for the NVSS sources, which could generate spurious power on large angular scales \citep{Smith_2007}.
The upper limit $\ell=310$ is set to reduce the impact of nonlinear effects in the matter power spectrum and, thus, to minimize deviations from the Gaussian statistics that we assumed in order to estimate the covariance matrix and perform the $\chi^2$ analysis. The large mismatch seen in the TGSS and NVSS auto-spectra at $\ell<40$ is not observed in the cross-spectra, which are instead in good agreement. 
Although this result does not clarify the origin of the large scale power excess detected in the TGSS auto-spectrum, it does support the hypothesis that it is not genuine but originates from observational systematics errors that could not be identified and corrected for. 
The absence of this excess in the cross-correlation validates the hypothesis that possible observational biases in the CMB lensing maps do not correlate with those that affect the TGSS catalog. Moreover, it confirms that cross-correlation analyses between  CMB lensing and catalogs of extra-galactic objects are less prone to observational systematic errors and, therefore, can safely be exploited to make inference on the nature, redshift distribution and clustering properties of the radio sources. We also stress that while the NVSS-CMB lensing cross-correlation signal has been already measured \citep{planck_lens2013}, this is the first time that such cross-correlation signal is detected using TGSS, which is a catalog that contains a different mixture of radio sources.

\item None of the combinations of the $b(z)+N(z)$ models proposed to fit the auto-power spectrum of the NVSS and TGSS sources, and that we  implemented in our analysis, are also able to fit the observed $\kappa g$ cross-spectrum over the full $\ell$-range. In this work we considered the two main existing extragalactic radio simulations publicly available, T-RECS and S$^3$, to model the redshift distribution $N(z)$ of the radio sources in the TGSS and NVSS catalogs. We did not consider their associated $b(z)$ model which basically reflects the bias of the dark matter halos host extracted from the parent N-body. Instead, we examined two bias models used in previous clustering analyses of the TGSS and NVSS objects (see \citealt{Nusser_2015,tiwari16,bengaly18,dolfi_2019}). The first one, dubbed HB, relies on the halo model applied to all types of radio sources in the T-RECS and S$^3$ simulations \citep{ferramacho2014}. The second one, PB relies on the physical model of \cite{Nusser_2015} and it assumes that only one galaxy can be hosted in a halo. 
At low redshifts direct observations provide some constraint on the bias of the various types of radio sources (i.e. \citealt{magliocchetti2017,Hale2018}).
On the contrary, at high redshift the bias of the radio sources largely relies on theoretical assumptions.
Both HB and PB predict that the bias steadily (and rapidly) increases with the redshift, which is somewhat nonphysical. For this reason, we also considered truncated versions of both models in which the bias amplitude of each radio source population is kept constant for $z\geq 1.5$. The choice of this redshift value is quite arbitrary. However, we verified that it has little impact on the results, which are quite insensitive to the choice of the truncation redshift.

To quantify the agreement between model and data, we performed a $\chi^2$ analysis.
To model the angular spectra, we assumed a flat $\Lambda$CDM Planck Cosmology \citep{Planck_2018} and the $N(z)$ and $b(z)$ models described above. The comparison has been done in bins of $\Delta \ell=30$ to compromise between the need to minimize the loss of information due to data compression and that of reducing the covariance among different bins. The latter requirement allows us to use an analytic diagonal Gaussian approximation to estimate the covariance matrix. The goodness of the Gaussian hypothesis, has been checked by comparing its results with those obtained by repeating the $\chi^2$ analysis using a numerical covariance matrix estimated with the Jackknife re-sampling method (Appendix \ref{sec:jk_cov}).

The results of the $\chi^2$ analysis indicate that no $N(z)$ and $b(z)$ models combination succeeds in fitting the cross-spectrum over the full multipole range. Most of the considered models provide a good fit to the data over much of the multipole range except in the first bin $[11,41]$ and at $\ell \sim 150$ where they underpredict the amplitude of the cross power spectrum. This is true for both the TGSS and the NVSS catalogs.
The power peak at $\ell \sim 150$ is a peculiar feature that has no counterpart in the auto-spectra which are dominated by shot noise at these multipoles.
The excess power in the first $\ell$-bin, $[11,41]$, is more striking.
A similar excess is also seen in the NVSS auto-spectrum on the same scale, as shown by \cite{dolfi_2019}, when it is compared to the HB + $S^3$ model.
We found that, not only HB + $S^3$ fails to match the cross-spectrum on the same scales, but other model combinations (PB + $S^3$ and PB + T-RECS) under-predict the NVSS and TGSS - CMB lensing cross power on large angular scales. Indeed, the only model combination that fits the angular cross spectra in the first bin is the HB + T-RECS. The reason for this is related to the rapid evolution of the bias with redshift, which boosts up the clustering amplitude of the objects at $z<1.5$, since similar results are also obtained with the TPB and THB models in which the bias of each source population is fixed to be constant beyond $z=1.5$. On the other hand, the HB + T-RECS model combination consistently overpredicts the power amplitude at all $\ell > 41$ except for the bin centered at $\ell \sim 150$ and provides a much worse fit to the data than the other models, as indicated by the PTE values.

\item
When the same bias model is assumed (like in the PB or TPB cases) the $\kappa g$ model is rather insensitive to the choice of the $N(z)$. 
This robustness of the CMB lensing cross-correlation to the $N(z)$ model uncertainties is a key feature that allows us to break the degeneracy between $b(z)$ and $N(z)$ when combined to auto-correlation measurements, as recently pointed out also by  \cite{Alonso_2021}. We therefore performed a joint auto- and cross- correlation analysis using the CMB lensing and the NVSS catalog only.
Unfortunately, the limited signal-to-noise of the auto-spectrum is not enough to further discriminate among the 
various combinations of $b(z)$ and $N(z)$ models. The analysis confirms that none of these combinations succeeds in 
fitting the auto- and cross-angular spectra on both large and small scales. 

\item Intriguingly, a power excess at  $\ell <41 $ has also been recently detected in the auto-spectrum of the radio sources in the RACS catalog by \cite{Bahr-kalus2022}. An excess that no effective bias model combined with either $S^3$ or T-RECS redshift distributions is able to match. Suspecting a spurious origin of the excess, \cite{Bahr-kalus2022} compares the results obtained when including or excluding multipoles $\ell <41 $ and find that in the latter case the minimum reduced $\chi^2$ decreases by a factor 3-10, depending on the model combination. When we repeat the same exercise and remove the first $\ell$ bin from our cross- and joint- angular spectra analyses, we also found that the 
reduced $\chi^2$ value decreases but only by 30-60\%, which hardly indicates a spurious origin of this large scale angular power.

\item We checked the robustness of our results to a number of  potential sources of systematic errors and found that they do not change when different flux cuts are used to select the radio samples or when different geometry masks are considered to account for the impact of Galactic foreground. The CMB lensing signal is also prone to potential biases. Therefore, we checked the sensitivity of our results to using
different Planck CMB lensing reconstructions as well as to the magnification bias modelling, and found no significant effect. Moreover, we cross-correlated the radio sources data maps with Planck CMB lensing simulations that include realistic reconstruction noise and we recovered a null mean cross-spectrum, thus confirming that possible spurious signals in the radio sources maps do not correlate with lensing (see Appendix \ref{sec:sims_cov}). Finally, the results of our $\chi^2$ analysis are potentially sensitive to the Gaussian hypothesis, that we have used to generate the covariance matrix of the auto- and cross-spectra, and to the choice of the $\ell-$ binning. These tests, that are described in the Appendix \ref{sec:cov_appendix}, also confirm the robustness of our results.

\item As none of the considered $b(z) + N(z)$ models can successfully fit the cross-spectra, and given the robustness of the model prediction to the uncertainties in the composition and redshift distribution of the radio sources, we repeated the analysis with a fixed $N(z)$ model (either T-RECS or $S^3$) and let a bias parameter free to vary. Specifically, we explored two simple bias models commonly used in the clustering analyses of the radio sources (see also \citealt{Alonso_2021,Bahr-kalus2022}): a constant bias and a bias which evolves with the inverse of the linear growth factor $D(z)^{-1}$. Both models depend on a single free parameter, $b_g$, the effective bias of the sample at $z=0$. Leaving the bias free to vary generally  reduces, but does not eliminate, the mismatch between the model and the measured cross-spectra in the first redshift bin, though it does not improve the quality of the fit in correspondence to the power peak at $\ell \sim 150$.
The constant bias model outperforms the more physically motivated redshift evolving one for both choices of $N(z)$ and for both the NVSS and TGSS cross-spectra.

Focusing on the NVSS case, we find that the best fit $b_g$ values obtained for the non evolving bias model (2.53 and 2.79 for T-RECS and $S^3$ case, respectively) are systematically larger than those of the $D^{-1}(z)$ evolving model (1.64 and 1.73).
The difference is significant compared to the typical 1-$\sigma$ uncertainty of 5 \% on $b_g$.
Since $b_g$ represents the effective bias of the NVSS sample at $z=0$, this large value would imply the presence of a local population of radio sources that are significantly biased with respect to the mass density field. This seems in contradiction with the evidence that at low redshift the NVSS sample is dominated by low-biased SFGs and FRI sources. Interestingly, a similar result has been found in the auto-correlation analysis of the RACS sample \citep{Bahr-kalus2022}. The RACS sample, like the NVSS one, is locally dominated by SFGs and FRI sources. However, the constant bias best fitting model requires a high linear bias parameter, 2.41 or 3.24, depending on the $N(z)$ model adopted.
Evidently, in their case as well as in ours, a large $b_g$ value is required to fit the power on large angular scales that is mainly contributed by the local large scale structure.
Invoking a bias that increases with redshift, either linearly or exponentially, significantly reduces the  best-fit $b_g$ values, like in our case. However, and unlike our case, the adoption of an evolving bias in the RACS analysis either improves or does not significantly modify the quality of the fit.

We ascribe the preference for a constant bias model with a large $b_g$ parameter to the attempt to fit the large scale power at $\ell < 40 $. 
Interestingly, \cite{Bahr-kalus2022} also find a large power excess in the auto-spectrum for $\ell < 40 $. An interval of multipoles that they need to exclude from the analysis to obtain reasonably good fit to the data. When they cross-correlate the positions of the RACS sources with the CMB temperature map this large scale power excess, which they interpret as spuriously generated by observational systematics, is largely removed. As it should be in our cross-correlation analysis. For this reason, we are very cautious in dismissing the cross-power in the first multipole bin as non-genuine.

We repeated the same analysis for the TGSS  - CMB lensing cross spectra. Results are consistent with those obtained in the NVSS case, except for the fact that all best fit $b_g$ values are systematically larger (by $\sim 10$ \%), which is expected given the brighter nature of the TGSS sources.

\item We performed a joint cross- and auto-correlation analysis to further test the bias models described above. Adding the auto-spectrum information does have an impact. Overall the results of the joint analysis confirm that of the cross-correlation only case, however, the addition of the auto-spectrum information 
indicates a stronger preference for a constant bias (with a very similar $b_g$ parameter) over the evolving bias. This result clearly confirms the importance of joining auto- and cross-correlation spectra, not only to identify and remove systematic uncertainties, but also to discriminate among competing models and break parameter degeneracy. 

\end{itemize}

To summarize: we have shown that the cross-correlation analysis with CMB lensing does not present an anomalous large scale power for the TGSS sample, hence confirming the spurious nature of the excess found in the auto-spectrum. However, the amplitude of both the NVSS and the TGSS cross-spectra in the multipole range $\ell=[11,40]$ remains high. None of the considered $\Lambda$CDM-based cross-spectra models that are based on physically motivated $b(z)$ and $N(z)$ of the radio sources succeed in matching that large scale cross-power.
There are three possible explanations to this failure.

The first one is that some systematic errors are still present in the cross-correlation analysis. This would only be possible if systematic errors that affect the radio catalogs correlate with those that may affect the CMB lensing maps. However, considering the very different nature of these two types of tracers and the possible sources of systematic errors that may affect them, this seems unlikely. Although, the presence of a small, but significant, excess power at $\ell \sim 150$ corresponding to a $\sim 1^{\circ}$ angular scale, may indicate the opposite.

The second possibility is that either the $N(z)$ or the $b(z)$ models are inadequate (or both).  Our analysis shows that, irrespective of the $N(z)$ model adopted, no biasing scheme is able to provide enough large scale power to match the observed cross-correlation amplitude. The best performing models require a rather nonphysical constant bias scheme characterized by a large linear bias value that seems to be inconsistent with that of the radio sources that populate the local Universe.
One possible way out is to advocate a bias model in which the effective bias parameter of radio sources is a decreasing, rather than increasing, function of the redshift. Such a model was indeed advocated by \cite{Hernandez_Monteagudo_2010}
to reproduce the auto-correlation of NVSS sources as well as their cross-correlation with the CMB temperature map measured by WMAP. To check this possibility, we considered an alternative model in which the bias evolves with the linear growth factor $b(z)\propto D(z)$, rather than its inverse. Results were not satisfactory, since the resulting effective bias parameter came out to be $b_{g}\simeq 3$. This is consistent with the idea of the \cite{negrello,Raccanelli2008} model according to which local AGN-powered radio sources are rare and reside within rich clusters of galaxies, however the quality of the fit did not improve.
Interestingly, \cite{Bahr-kalus2022} have pointed out that the function $N(z)\times b(z)$ that best fits their auto and cross-spectra peaks at $z\sim 1$, which again, seems to disfavor the case of an increasing $b(z)$ model, if T-RECS and $S^3$ predictions are robust. We conclude that the problem of determining the composition, redshift distribution and clustering properties of the continuum radio surveys is an open one that will likely be solved by combining auto-and cross-correlation analyses like the one presented here and by increasing the fraction of radio objects with measured redshift by means of dedicated observational campaigns.

The final possibility, of course, is that the $\Lambda$CDM model is incorrect. We do not wish to insist on this tantalizing possibility, since our analysis does not provide unique evidence to point along this direction. However, we wish to stress that the excess power we have discussed so far is on angular scales of $\sim 10^{\circ}$, so it is not obviously related to the much discussed radio dipole excess.

\begin{acknowledgements}
We thank Tommaso Giannantonio, Giulio Fabbian and Nicola Vittorio for useful discussions, Isabella Prandoni for help with the modelling of radio sources number counts, and Antony Lewis for support with the CAMB code. Enzo Branchini is partly supported by ASI/INAF agreement n. 2018-23-HH. “Scientific activity for Euclid mission, Phase D", by ASI/INAF agreement n. 2017-14-H.O “Unveiling Dark Matter and Missing Baryons in the high-energy sky" and by
MIUR/PRIN 2017 “From Darklight
to Dark Matter: understanding the galaxy-matter connection to measure the Universe".
Enzo Branchini, Marina Migliaccio and Giulia Piccirilli are also supported by the INFN project “InDark". Marina Migliaccio acknowledges partial support by ASI/LiteBIRD grant n. 2020-9-HH.0. 
\end{acknowledgements}

\appendix
\counterwithin{figure}{section} 
\section{Shot noise estimation}
\label{sec:shot_noise_estimation}
When dealing with the $gg$, auto-spectrum we need to account for a constant noise term, $N^{gg}$, that includes the effect of Poisson noise and spurious contribution from multiple sources. The Poisson noise term can be estimated from the mean number density of the sources while the multiple source contribution can be inferred by fitting a 1-halo term to the auto-correlation function of the sources at small angular separations \citep{Blake_Wall_2002}. These noise terms have been 
estimated by \cite{dolfi_2019} for TGSS and NVSS samples similar to ours. Once $N^{gg}$ is subtracted from the measurements, the estimated power spectra should approach zero at large $\ell$ values, where the noise dominates.
This indeed occurs in the TGSS case. On the contrary, 
after subtracting the $N^{gg}$ terms estimated by \cite{dolfi_2019}, the NVSS auto-spectrum exhibits negative residuals at high multipoles.

Therefore, we decided to follow a different procedure and estimate the NVSS noise term by enforcing the angular spectrum to approach zero at high $\ell$ values, where it flattens out. We do so in Section~\ref{sec:auto_spect} by minimizing the $\chi^2$ between the measured and the model auto-spectrum in which a free $N^{gg}$ term is added to the latter. The result of this procedure depends on the model auto-spectrum. To check the sensitivity of the noise correction to  the choice of $N(z)$ and $b(z) $ we repeated the procedure for all the model combinations considered in our analysis. 

The results are summarized in Table~\ref{tab:shot_values}. The best fit values obtained from the auto-spectrum analysis are indicated as $(N^{gg})^a$ in the second and fourth columns and they depend on the assumed model combination. In all cases uncertainties (not indicated in the table) are of the order of $2\%$. 


We replicated the $\chi^2$ minimization also in the joint analysis, where both $gg$ and $\kappa g$ spectra are considered. The resulting noise terms are indicated as 
$(N^{gg})^j$ in Table~\ref{tab:shot_values}. They are almost identical to those obtained from the auto-spectra only. 
Note that these are the noise values that have been subtracted to the auto spectra in the joint analysis presented in the main text. We stress the fact that these $N^{gg}$ values are typically smaller than
the one estimated by \cite{dolfi_2019} i.e. $2.07\times10^{-5}$. 

\begin{table}[H]
\centering
\begin{tabular}[t]{l  r r | r r }
\hline
\hline\xrowht[()]{5pt}
Bias&$(N^{gg})^a_{S^3}$&$(N^{gg})^j_{S^3}$&$(N^{gg})^a_{T-RECS}$&$(N^{gg})^j_{T-RECS}$\cr
\hline
HB& $1.92\times 10^{-5}$ & $1.93\times 10^{-5}$ &  $1.70\times 10^{-5}$&  $1.70\times 10^{-5}$\cr
THB&  $1.99\times 10^{-5}$ & $1.98\times 10^{-5}$ &  $1.70\times 10^{-5}$&  $1.70\times 10^{-5}$\cr 
PB&  $1.92\times 10^{-5}$ &  $1.92\times 10^{-5}$ & $1.88\times 10^{-5}$&  $1.89\times 10^{-5}$\cr
TPB &   $1.95\times 10^{-5}$ &  $1.94\times 10^{-5}$ & $1.89\times 10^{-5}$ &  $1.89\times 10^{-5}$\cr 
 \hline
 \end{tabular}
\caption{Values of the NVSS auto-spectrum noise correction $N^{gg}$ estimated for various model combinations and for both the auto-spectrum only analysis (second and fourth columns, superscript $^a$) and joint analysis (third and fifth columns, superscript $^j$).
The values listed in the different rows are obtained with different bias models, indicated in column 1. Results obtained with the $S^3$ model are grouped in the left part of the Table. Those obtained with T-RECS model are grouped in the right part.
}
\label{tab:shot_values}
\end{table}


\section{Covariance matrix estimation}
\label{sec:cov_appendix}
\subsection{The Jackknife method}
\label{sec:jk_cov}
All the $\chi^2$ analyses presented in the main text, have used analytical expressions for the covariance matrices that rely on the hypothesis of Gaussian statistics. To guarantee the validity of this hypothesis and reduce the covariance among the $\ell-$ modes, we decide to bin the measured and the modelled angular spectra using a rather large $\ell$-bin, $\Delta_{\ell} = 30$.
The scope of this Appendix is to check the goodness of this strategy and justify the choice of the bin size.

To check the adequacy of the Gaussian hypothesis, we compare the theoretical covariance matrix for the NVSS - CMB lensing cross-correlation spectrum specified in Equations \ref{eq:cov_joint} and \ref{eq:diag_cov}, with an alternative numerical evaluation based on the Jackknife (JK hereafter) resampling method.
The JK method allows us to estimate the errors and their covariance from the data without relying on theoretical modeling.  
To evaluate the JK matrix we follow the procedure outlined in \cite{Norberg_2009}:
\begin{enumerate}
    \item 
    We use \mintinline{Python}{HEALPix} with  $N_{side} = 4$ to
    divide the  NVSS and CMB-lensing maps into $N_{sub}=192$ independent, equal area patches. We only keep the 97
    patches whose overlap with the unmasked region is more than $20\%$.
    \item We create a pair of NVSS and CMB-lensing maps in which one of the 97 patches has been removed.
    \item 
    We estimate the cross-spectrum of this i-th map pair $\hat{C}^{\kappa g, i}_{\ell}$ 
    using Anafast\footnote{\url{https://healpix.sourceforge.io/html/fac_anafast.htm}}.
    \item 
    We go back to step 1, unless $N_{tot}=N_{sub}=97$ pairs of maps have been generated and their cross-spectra have been computed. In that case we proceed to the next step.
    \item
    We estimate the JK covariance matrix as 
    \begin{equation}
\label{eq:jk_cov}
    \mathcal{C}ov^{JK}_{\ell \ell'} = \frac{N_{sub} -1}{N_{sub}} \sum_i \left(\hat{C}^{\kappa g, i}_{\ell}- \bar{C}^{\kappa g}_{\ell}\right) \left(\hat{C}^{\kappa g, i}_{\ell'}- \bar{C}^{\kappa g}_{\ell'}\right),
\end{equation}
where the sum runs over all $N_{sub}$ map pairs, $\bar{C}_{\ell}^{\kappa g} = \dfrac{1}{N_{sub}}\sum_i \hat{C}_{\ell}^{\kappa g, i}$ and the multiplicative factor $N_{sub} -1$
accounts for the fact that Jackknife resampling generates correlated data-sets.
\item We estimate the normalized correlation coefficient matrix:
\begin{equation}
\label{eq:coerr_matrix}
    R_{\ell \ell'}=\frac{\mathcal{C}ov_{\ell \ell'}^{JK}}{\sqrt{\mathcal{C}ov^{JK}_{\ell \ell} \mathcal{C}ov^{JK}_{\ell' \ell'}}}.
\end{equation}
\end{enumerate}

\begin{figure}
\centering
\includegraphics[width=0.85\hsize]{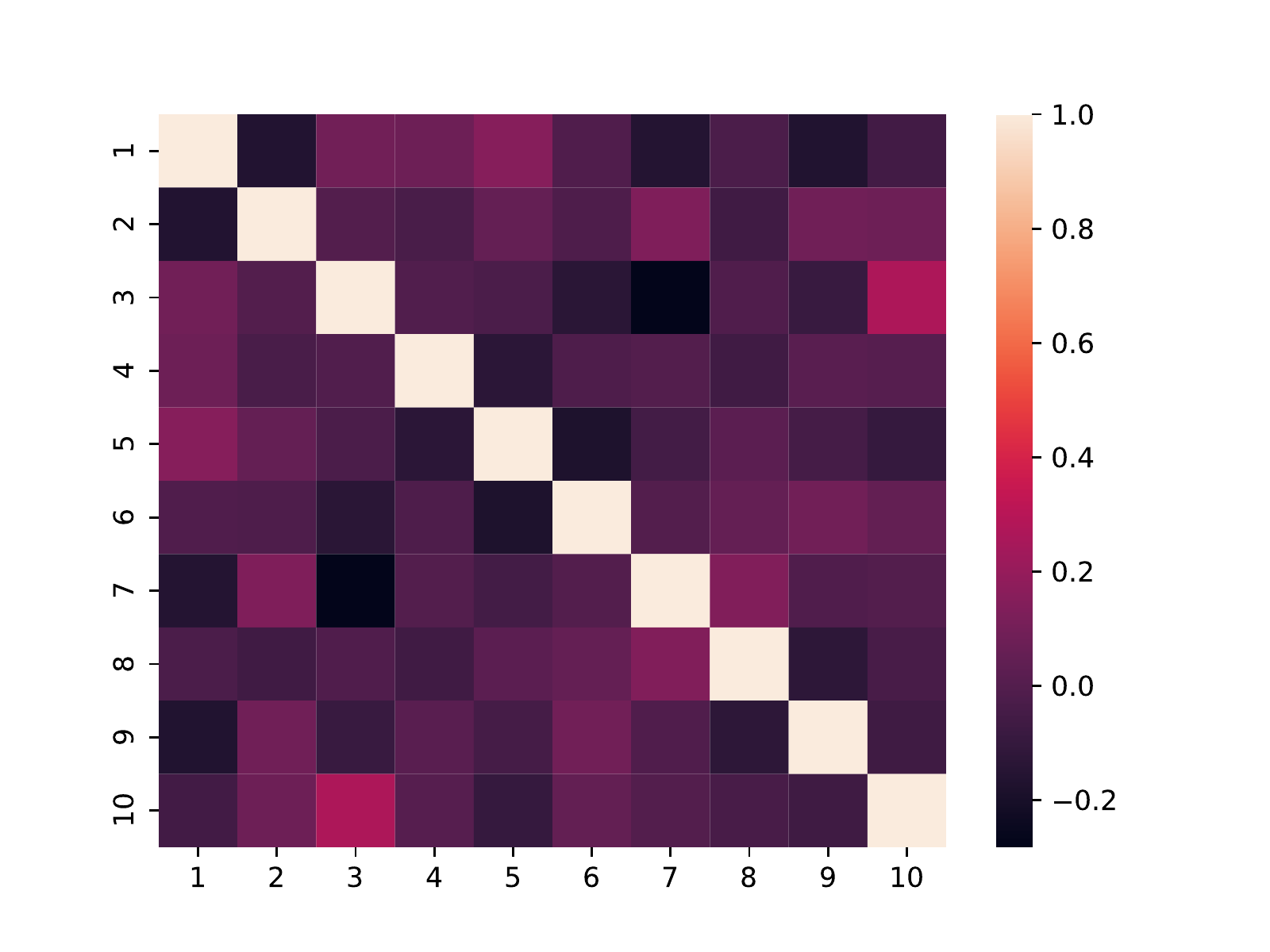}
\caption{ $10\times 10$ JK correlation coefficient matrix $R_{\ell \ell'}$ for the $\kappa g$ NVSS - CMB lensing cross spectrum evaluated in bins of size $\Delta_{\ell}=30$. The 
value of the off-diagonal cross-correlation coefficients is color-coded according to the vertical bar.}
\label{fig:jk_cov}
\end{figure}

The correlation coefficient matrix of the $ \kappa g$ cross-spectrum evaluated in 10 equally spaced 
$\Delta_{\ell}=30$ bins in the range $11\leq \ell \leq 310$
is shown in Figure~\ref{fig:jk_cov}.  The visual inspection reveals that the amplitude of the off-diagonal elements is generally small,
which is an indication that the size of the $\Delta_{\ell}$ bin is effectively suppressing the error covariance.

\begin{figure}[H]
\centering
\includegraphics[width=\hsize]{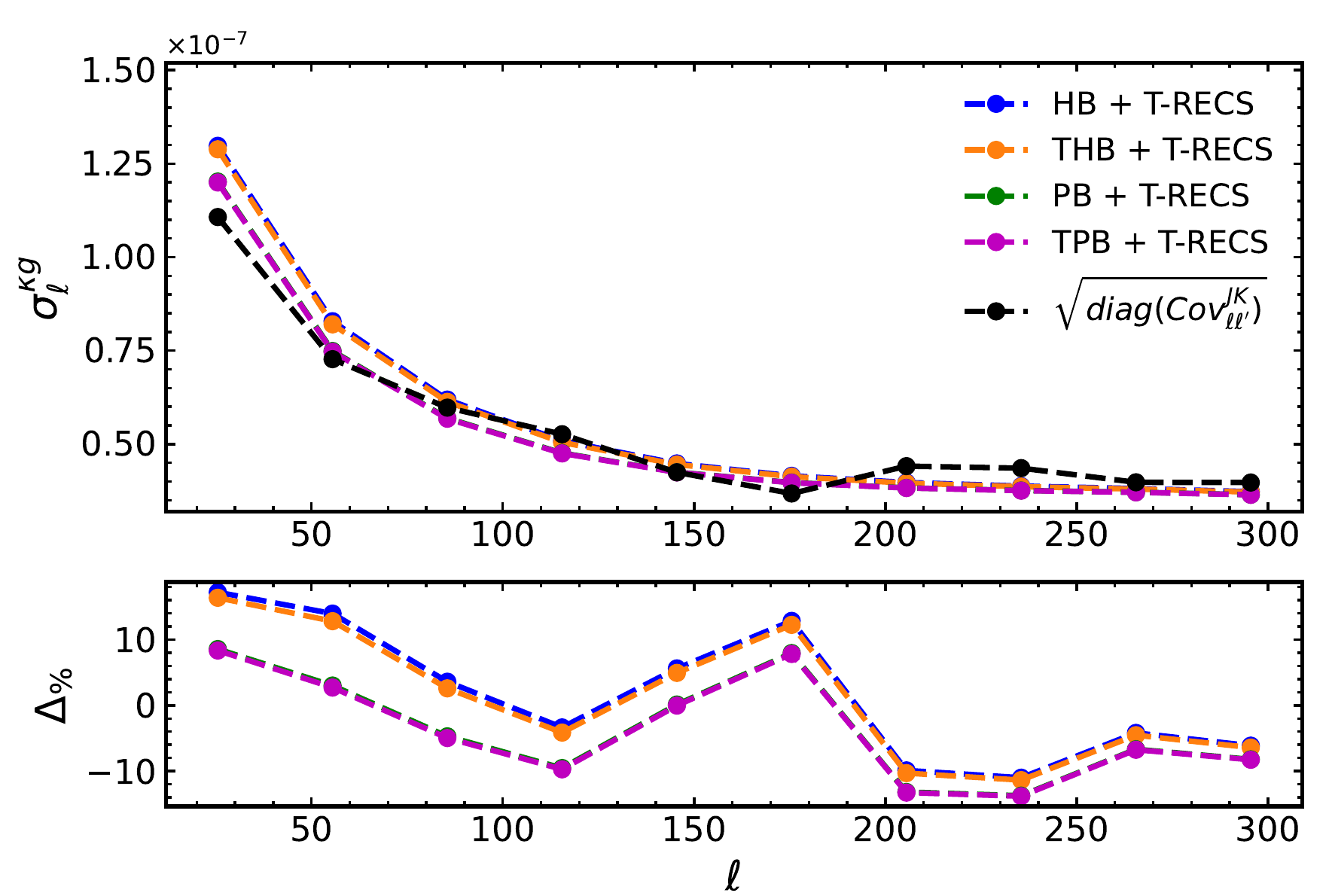}
\caption{
Gaussian vs. JK errors for the 
$\kappa g$ NVSS cross spectrum estimated in bins $\Delta_{\ell}=30$ in the range $11\leq \ell \leq 310$.
Upper panel. Colored dots and dashed curves indicate Gaussian errors for different $b(z)+N(z)$ model combinations specified in the labels. Black dots and dashed curve: JK errors.
Errors have been estimated from the diagonal elements of the Gaussian covariance (Equation~\ref{eq:diag_cov}) and JK covariance Equation~\ref{eq:jk_cov}.
Bottom panel: per cent difference between Gaussian and JK errors.}
\label{fig:sigma2_cov}
\end{figure}

To make this statement  more quantitative and assess the goodness of the Gaussian model, we compare in Figure~\ref{fig:sigma2_cov}
the diagonal elements of the Gaussian covariance matrix of the cross-spectrum (curves with different colors) with the same elements in the JK covariance matrix (black dots).
For the Gaussian cases we plot, in the upper panel, different curves corresponding to the different $b(z)+N(z)$ model combinations assumed to estimate the covariance. Their similarity indicates that Gaussian errors are not very sensitive to the $N(z)$ and $b(z)$ choice.
The bottom panel quantifies the percentage difference between the Gaussian errors and the JK errors.
The differences between the two types of errors oscillate with $\ell$
with a modest amplitude of the order of 10\%. Moreover, the fact that these oscillations are of the same order of those seen in the black curve around its smoothed interpolation suggest that uncertainties in the estimate of the JK errors are of the same order of the differences between JK and Gaussian estimates, i.e. per cent discrepancies in the bottom panel are not solely driven by deviations from the Gaussian hypothesis.
We also notice that differences among Gaussian errors obtained with various 
$N(z)$ and $b(z)$ schemes steadily decrease with $\ell$.

In our analysis, we considered binned spectra and we chose a bin size $\Delta_{\ell} = 30$. As anticipated, we set the size wide enough to guarantee that the hypothesis of Gaussian statistics would hold and that a Gaussian model could have been adopted to perform the $\chi^2$ analysis. 
To justify this choice, we repeated the analysis using different bin sizes $\Delta_{\ell} = 20, 30$ and $50$ and compared the results obtained using a Gaussian vs. a JK covariance matrix.
The idea is that the two sets of results need to be consistent, and a smaller $\chi^2_{min}$  would be obtained for the bin size $\Delta_{\ell}$ that minimizes the error covariance while preserving the information.

Since the results are not very sensitive to the choice of $N(z)$ and $b(z)$,
in Table~\ref{tab:chi2_jk} we only show the ones obtained for the combination $S^3$ and TPB.
When the Gaussian errors are used (second column), the $\chi^2_{min}$ value steadily increase with the bin size (and, correspondingly, the PTE decreases). 
We interpret this result as an evidence that, when the number of bins is reduced, the statistical significance of the mismatch at small $\ell$ values increases.
The case with the JK errors (third column) is different. The dependence of $\chi^2_{min}$ value on $\Delta_{\ell}$ is not monotonic. There is a soft spot at $\Delta_{\ell} = 30$ in correspondence of  which the reduced $\chi^2_{min}$ value is similar to the one obtained in the Gaussian case.
Moreover, for $\Delta_{\ell} = 20$ the reduced $\chi^2_{min}$ value of the Gaussian analysis, which ignores error covariance altogether, is smaller than the JK one. This suggests that with this bin size the Gaussian errors could be overestimated.
Therefore, we decided to make a conservative choice and adopt, for the analysis presented in this work, $\Delta_{\ell} = 30$.

The adequacy of this choice is corroborated by the results obtained with other $N(z)$ and $b(z)$ models. Adopting, for example, the T-RECS + TPB combination we find reduced $\chi^2_{min}$ values that are on average 30 \% smaller than those obtained with the $S^3$ case. However the $\chi^2_{min} -\Delta_{\ell}$ behavior is the same, including the existence of a soft spot at $\Delta_{\ell} = 30$ for the JK case and the corresponding similarity between the $\chi^2_{min}$ values of the Gaussian and JK analyses.

\begin{table}
\centering
\begin{tabular}[H]{l  c  c}
\hline
\hline\xrowht[()]{5pt}
$\Delta_{\ell}$&$\chi^{2}_{ANAFAST}$/d.o.f.(PTE)&$\chi^{2}_{JK}$/d.o.f.(PTE)\cr
\hline
20  &  1.20 ($2.6\times 10^{-1}$)   &  1.59 ($6.7 \times 10^{-2}$) \cr 
30  &  1.75 ($6.4\times 10^{-2}$)    & 1.51 ($1.3\times 10^{-1}$) \cr 
50  &  2.35 ($2.9\times 10^{-2}$)   & 1.99  ($6.3\times 10^{-2}$) \cr 
 \hline
 \end{tabular}
 \caption{ Reduced $\chi^2_{min}$ values (and corresponding PTEs) obtained for the analysis of the $\kappa g$ cross spectrum measured in bins of different sizes $\Delta_{\ell}$, (column 1) in the range $\ell = [11,310]$.
 The same model combination T-RECS and TPB has been adopted. Column 2: results obtained assuming Gaussian errors. Column 3: 
 results obtained assuming JK errors. }
\label{tab:chi2_jk}
\end{table}

\subsection{Simulation based tests}
\label{sec:sims_cov}

In this appendix, we provide a further validation of the analytic Gaussian approximation adopted for the power spectrum covariance matrix by performing two different sets of tests based on simulations. As a by-product, the tests also demonstrate that our pipeline is unbiased.

\begin{figure}[!t]
\centering
\includegraphics[width=0.98\hsize]{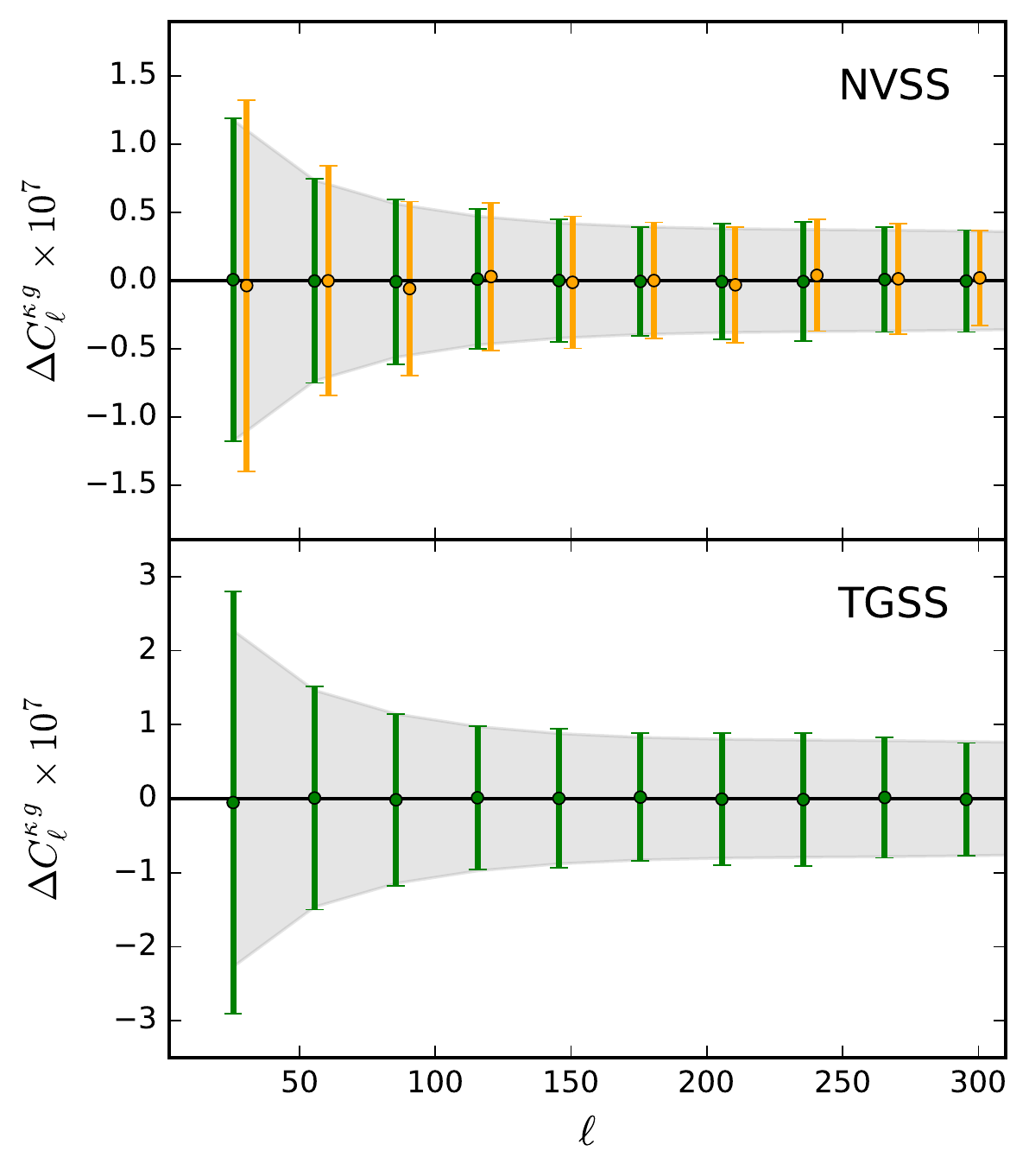}
\caption{Mean recovered cross-power spectra from realistic simulations of Planck CMB lensing convergence and NVSS or TGSS data maps (in green). The spectra are consistent with zero, demonstrating that our analysis pipeline does not induce spurious cross-power in the absence of correlation (see text). Error bars are the diagonal elements of the empirical covariance matrix derived from the same simulations. These need to be compared with the grey bands that represent the analytic Gaussian uncertainties used throughout the paper. In the top panel, in orange, we also show the mean cross-power spectrum from Gaussian correlated lensing and galaxy simulations minus the input theoretical cross-power spectrum. Consistency with zero implies that our pipeline is able to recover in an unbiased way a known input cross-spectrum. Error bars in this case are the diagonal elements of the empirical covariance matrix derived from the correlated simulations.}
\label{fig:sigma_cov_sims}
\end{figure}

\begin{itemize}
    \item As a first approach, we cross-correlate $300$ realistic simulations of the Planck CMB lensing convergence\footnote{\url{https://wiki.cosmos.esa.int/planck-legacy-archive/index.php/Lensing}} \citep{planck_lens} with the radio source maps of both NVSS and TGSS. Each simulation contains a realization of the CMB lensing convergence, drawn from the Planck best-fit power spectrum, plus realistic lensing reconstruction noise, which also properly accounts for the fact that the reconstruction has been performed on a masked sky. When extracting the cross-angular power spectra we employ the same masks used in our main analysis. Then, the covariance matrix is estimated from the simulated cross-spectra. However, lensing simulations and radio source data maps are in principle uncorrelated, unless there are anomalous features in the latter that can produce spurious correlations. As a consequence, this procedure does not account for the cosmic variance contribution to the uncertainties that should come from the correlated part of the maps, $C^{\kappa g}_{\ell}$. Regardless, we expect the estimated covariance to be fairly representative as both lensing and galaxy maps are noise dominated at the relevant scales used in our analysis. In Figure \ref{fig:sigma_cov_sims} we show the mean recovered cross-spectrum from simulations and the $1\,\sigma$ errors obtained as the standard deviation of the simulated spectra. Finding that the mean cross-spectrum
    is consistent with zero serves as a powerful check that: a) our analysis pipeline does not induce any spurious bias; b) the large-scale anomalous power in TGSS does not correlate with CMB lensing and its reconstruction noise. The figure also provides a comparison to the analytic Gaussian uncertainties used in our main analysis, which are shown with the grey band. We can appreciate a good agreement between the two, with the only exception of the uncertainty in the first bin for TGSS. In this case the uncertainty from simulations is larger by $\sim 25\%$, this is a consequence of the excess power at large scales in the TGSS clustering which is not reproduced by any theoretical model. Nevertheless, we have verified that this difference has negligible impact on the $\chi^2$ analysis and does not alter the conclusions of our work. Finally, checking the off-diagonal elements of the covariance matrix estimated from simulations, we have verified that bin-to-bin correlations are $<20\%$ everywhere. 
    \item In the second approach, starting from the theoretical spectra of the model corresponding to the combination T-RECS + TPB, we simulate $300$ maps of both the galaxy density contrast and the CMB lensing convergence as Gaussian correlated fields, this time, with the expected level of correlation $C^{\kappa g}_{\ell}$ (see e.g. Appendix of \citealt{Giannantonio2008}). We include the contribution of noise as well, which for the lensing, in this case, has been generated as simple Gaussian white noise from the $N^{\kappa \kappa}_{\ell}$ power spectrum, while for the galaxy maps consists in Poisson realizations with the shot noise level $N^{gg}$. We estimate the cross-spectra of the simulated maps after applying the same masks used in the main analysis. We only consider the NVSS case, as TGSS will provide similar results given the mask for this catalog is very close to that of NVSS. From Figure \ref{fig:sigma_cov_sims}, we can see that our pipeline is able to recover the correct cross-correlation spectrum from the simulations. Moreover, error bars estimated from simulations are in good agreement with the analytic Gaussian uncertainties with differences of at most $15\%$.
    
\end{itemize}

\bibliographystyle{aa} 
\bibliography{bibliography}
\end{document}